\documentclass[prb,twocolumn,showpacs,floatfix]{revtex4}

\usepackage{amssymb,amsmath}
\usepackage[mathscr]{eucal}

\usepackage{graphicx}

\begin{document}

\title{Conduction eigenchannels of atomic-sized contacts:\\
$\boldsymbol{Ab\ initio}$ KKR Green's function formalism. }

\author{
Alexei Bagrets$^{1,2}$, Nikos Papanikolaou$^{3}$, and Ingrid Mertig$^{1}$ }

\affiliation{
$^1$Martin-Luther-Universit\"at Halle-Wittenberg, Fachbereich Physik, D-06099 Halle, Germany \\
$^2$Institut f\"{u}r Nanotechnologie, Forschungszentrum Karlsruhe, D-76344, Germany \\
$^3$Institute of Microelectronics, NCSR "Demokritos", GR-15310 Athens, Greece }

\date{\today}

\begin{abstract}
We develop a formalism for the evaluation of conduction eigenchannels
of atomic-sized contacts from first-principles. The multiple scattering
Korringa-Kohn-Rostoker (KKR) Green's function method is combined
with the Kubo linear response theory.
Solutions of the eigenvalue problem for the transmission matrix
are proven to be identical to eigenchannels introduced by Landauer and B{\"u}ttiker.
Applications of the method are presented by studying ballistic electron
transport through Cu, Pd, Ni and Co single-atom contacts.
We show in detail how the eigenchannels are classified in terms of
irreducible representations of the symmetry group of the system as well
as by orbital contributions when the channels wave functions are projected
on the contact atom.
\end{abstract}

\pacs{73.63.Rt, 73.23.Ad, 75.47.Jn, 73.40.Cg}

\maketitle

\section{Introduction}

The invention of the scanning tunneling microscope\cite{STM}
in 1981 and a consequent development in the beginning of the nineties
of the remarkably simple experimental technique known
as mechanically controllable break junction (MCBJ)\cite{Muller,MCBJ}
led to the possibility of fabrication of metallic point
contacts approaching the atomic scale. The recent review article
(Ref.~\onlinecite{Ruitenbeek}) summarizes the numerous achievements in
this field. In the experiments the conductance measured as a function
of the elongation of the nanocontacts decreases in a stepwise
fashion \cite{MCBJ,Ohnishi,Yanson_Nature_1,Cuevas_exp}
with steps of order of the conductance quantum $G_0 = 2e^2/h$.
Such behavior of the conductance is attributed to atomic rearrangements
that entails a discrete variation of the contact diameter.
\cite{Atom_rearrang,Brandbyge,Todorov,Agrait}

The electron transport in metallic nanocontacts
is purely ballistic and phase-coherent because their size
is much smaller than all scattering lengths of the system.
According to Landauer,\cite{Landauer}
conductance is understood as transport through nonmixing channels,
\[
 G = \frac{2e^2}{h} \sum_{n=1}^{N} T_n,
\]
where $T_n$'s are transmission probabilities. They are
defined as eigenvalues of the transmission matrix $\mathbf{\tau\tau}^{\dagger}$.
Here the matrix element $\tau_{nm}$ gives the probability amplitude
for an incoming electron wave in the transverse mode (channel) $n$ on the left
from the contact to be transmitted to the outgoing wave in the mode $m$ on the right.
Consequently, the eigenvectors of  $\mathbf{\tau\tau}^{\dagger}$ are
usually called eigenchannels.
It was shown in the pioneering work by Scheer {\it et al.}\cite{Scheer}
that a study of the current-voltage relation for the superconducting
atomic-sized contacts allowed to obtain transmission probabilities
$T_n$'s for particular atomic configurations realized in MCBJ experiments.
The $T_n$'s are found by fitting theoretical and
experimental $I-V$ curve which has a peculiar nonlinear behavior
for superconducting contacts at voltages $eV$ smaller than
the energy gap $2\Delta$ of a superconductor\cite{Scheer}.
The origin of such effect is explained in terms of multiple Andreev reflections.\cite{MAR}
The analysis of MCBJ experiments within the tight-binding (TB) model
suggested by Cuevas {\it et al.}\cite{Cuevas_TB_model,Scheer_Nature} gave a strong
evidence to the relation between the number of conducting modes
and the number of valence orbitals of a contact atom.

To describe the electronic and transport properties of nanocontacts,
quite a big number of different methods which supplemented each other
were developed during the last 15 years. Early models employed
a free-electron-like approximation.\cite{Brandbyge,FE_models,FE_Brandbyge}
Further approaches based on density functional theory (DFT) used psuedopotentials
to describe atomic chains suspended between jellium electrodes.\cite{Lang,Kobayashi}
The TB models were applied to the problem of the conduction
eigenchannels\cite{Cuevas_TB_model,TB_models} and to the study of the breaking
processes of nanowires.\cite{TB_models_1}
The up-to-date fully self-consistent
{\it ab initio} methods\cite{Ab-initio,Ab-initio_1,Rocha}
allowed to treat both the leads and the constriction
region on the same footing and to evaluate the non-equilibrium transport
properties as well\cite{Ab-initio_1,Rocha,new_methods}.

The scattering waves, underlying a concept of
eigenchannels introduced by Landauer and B{\"u}ttiker,\cite{Landauer}
do not form an appropriate basis for the most of {\it ab initio}\ methods.
Instead, one considers conduction channels as eigenvectors of some hermitian
transmission matrix written in terms of local,
atom centered basis set.\cite{Cuevas_TB_model,Ab-initio_1}
One of the goals of the present paper is to establish
a missing link between these approaches.
Below we introduce a formalism for the evaluation of conduction
eigenchannels, which combines an {\it ab initio}\ Korringa-Kohn-Rostoker
(KKR) Green's function method\cite{SKKR} for the electronic
structure calculations and the Baranger and Stone formulation of the
ballistic transport\cite{BarStone}. In recent publications,\cite{Imp_Paper1,Imp_Paper2}
we have successfully applied this method to the study of the electron transport
through atomic contacts contaminated by impurities.
In the present paper, mathematical aspects of the problem are considered,
followed by some applications. In particular,
we analyze the symmetry of channels and relate
our approach to the orbital classification of eigenmodes
introduced by Cuevas {\it et al.}\cite{Cuevas_TB_model}

The paper is organized as follows. A short description of the KKR
method is given in Sec.~II. We proceed in Sec.~III with a formal
definition of eigenchannels for the case of realistic crystalline
leads attached to atomic constriction. Sec.~IV supplemented by
Appendices~A and~B contains mathematical formulation of the
method. Briefly, using the equivalence of the Kubo and Landauer
approaches for the conductance, \cite{BarStone,Mavropoulos} we
build the transmission matrix $\tau\tau^{\dagger}$ in the
scattering wave representation. The angular momentum expansion of
the scattering Bloch states within each cell is used further to
find an equivalent, KKR representation of the transmission
operator for which the eigenvalue problem can be solved.
Applications of the method are presented in Sec.~V. In particular,
we focus on transition metal contacts (such as Ni, Co and Pd),
since experimental\cite{Ludoph,Half_int_exp,Viret} and theoretical
studies \cite{PRB_MR,TB-LMTO-Ni,Ni_Spain,Pt_Spain,Ni_Smogunov} of
their transport properties have been attracting much attention
during the last years. Experiments
\cite{Huge_MR,Huge_MR_1,Hua_Chopra,Sullivan_Ni,Chopra_Co}
regarding ballistic magnetoresistance (BMR) effect in
ferromagnetic contacts are commented. A summary of our results is
given in Sec.~VI.

%\clearpage

\section{Electronic structure calculation of the atomic contacts}

The systems under consideration consist of two semi-infinite crystalline
leads, left (L) and right (R), coupled through a cluster of atoms which
models an atomic constriction. In Fig.~1 a typical configuration
used in the calculations is shown --- the two
fcc (001) pyramids attached to the electrodes are joined via the vertex atoms.
We employed the {\it ab initio} screened KKR Green's function method
to calculate the electronic structure of the systems.
Since details of the approach can be found elsewhere, \cite{SKKR}
only a brief description is given below.

In the KKR formalism one divides the whole space into non-overlapping,
space-filling cells, with the atoms (and empty spheres) positioned at the sites ${\mathbf R}_n$,
so that the crystal potential $V$ is expressed in each cell
as $V_n({\mathbf r}) = V({\mathbf R}_n + {\mathbf r})$.
The one-electron retarded Green's function is
expressed in terms of local functions centered
at sites~${\mathbf R}_n$:
\begin{eqnarray}
\nonumber
\lefteqn{G^+({\mathbf R}_n +\ {\mathbf r},
{\bf R}_{n'} +\ {\mathbf r'};E) } \\
\label{Grr}
& = & \delta_{nn'}\sqrt{E} \sum_L R_L^n({\bf r}_<;E)H_L^n({\bf r}_>;E) \\
\nonumber
&  &\  +\ \sum_{LL'}R_L^n({\bf r};E)G_{LL'}^{nn'}(E)R_{L'}^{n'}({\bf r'};E)
\end{eqnarray}
where ${\mathbf r}$, ${\mathbf r'}$ are restricted to the cells
$n$ and $n'$; ${\mathbf r}_<$, ${\mathbf r}_>$ denote one of
the two vectors ${\mathbf r}$ or ${\mathbf r'}$ with the smaller
or the larger absolute value, and local functions
$R_L^n({\bf r};E)$ and $H_L^n({\bf r};E)$ are
the regular and irregular solutions of the
Schr\"odinger equation for the single potential $V_n({\mathbf r})$.
Here the index $L = (l,m)$ stands
for the angular momentum quantum numbers and atomic units are used:
$e = -\sqrt{2}$, $\hbar = 1$, $m = 1/2$.
The structural Green's function $G_{LL'}^{nn'}(E)$  (structure constants)
in Eq.~(\ref{Grr}) is related to the known structure
constants of the appropriately chosen reference system by the
algebraic Dyson equation which includes the difference
$\boldsymbol{\Delta t} =
\delta_{nn'} \delta_{LL'} \Delta t^{n}_{L}  $
between local $t$-matrices of the physical and a reference
system. In the screened KKR method\cite{SKKR}
we use a lattice of strongly repulsive,
constant muffin-tin potentials (typically, $\sim 4$Ry height)
as reference system that leads to structure
constants which decay exponentially in real space.

 Within the screened KKR method both a constriction region
and the leads are treated on the same footing.
This is achieved by using the hierarchy of Green's functions
connected by a Dyson equation, so that we perform
the self-consistent electronic structure
calculations of complicated systems in a step-like manner.
First, using the concept of principal layers
together with the decimation technique,\cite{2D_SKKR}
we calculate the structural Green's function
of the auxiliary system consisting of semi-infinite leads
separated by a vacuum barrier. At the second step,
the self-consistent solution of the impurity problem is found
by embedding a cluster with perturbed potentials
caused by the atomic contact into the auxiliary system.
Due to effective screening of perturbation,
the algebraic Dyson equation for the structure
constants is solved in real space.\cite{SKKR}

\begin{figure}[t]
\begin{center}
\includegraphics[scale = 0.29]{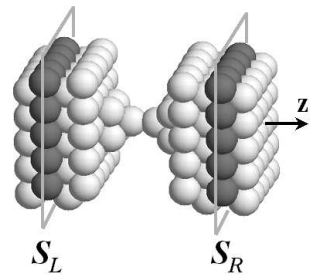}
\caption{Geometry of an atomic constriction:
two fcc (001) pyramids are attached via apex atoms.
Conductance is calculated between the two planes
$S_{\mathrm{L}}$ and $S_{\mathrm{R}}$ positioned in the leads.}
\end{center}
\end{figure}

\section{Definition of eigenchannels}

The concept of eigenchannels is introduced in the
Landauer approach to ballistic transport, where the problem
of the conductance evaluation is considered from viewpoint of
scattering theory. Following Landauer,\cite{Landauer}
we look at the system shown in Fig.~1 as consisting of
two semi-infinite leads (electrodes) attached to a scattering
region (atomic-sized constriction).
Far away from the scattering region the propagating states are the
unperturbed Bloch waves $\Psi^{\circ}_{\boldsymbol{k}}({\bf r},E)$
of the left (L, $z \to -\infty$) and right (R, $z \to +\infty$) leads,
where ${\boldsymbol{k}}$ belongs to the isoenergetic surface
$E = \mbox{const}$ (Fermi surface, $E=E_F$, in case of conductance)
and a common notation $\boldsymbol{k} = (\mathbf{k},\lambda)$
is used to denote Bloch vector $\mathbf{k}$ and band index $\lambda$.
For the eigenchannel problem one considers in-coming and out-going
states in the L and R leads normalized to a unit flux. The
in-states in L and out-states in R are
${\Phi}^{\circ}_{\boldsymbol{k}} =
{\Psi}^{\circ}_{\boldsymbol{k}}/\sqrt{v_{\boldsymbol{k}}}$
with positive velocity $v_{\boldsymbol{k}} \propto
\mbox{v}^z_{\boldsymbol{k}} > 0$ along $z$-axis.
The conjugated states $\Phi^{\circ\,*}_{\boldsymbol{k}} =
\Psi^{\circ\,*}_{\boldsymbol{k}}/%
\sqrt{|v_{\boldsymbol{-k}}|}$ are the out-states in L and
in-states in R with negative velocity $v_{\boldsymbol{-k}}
\propto \mbox{v}^z_{-\boldsymbol{k}} < 0$.
Here $\mbox{v}^z_{\boldsymbol{k}}$ is a $z$-component of the group
velocity  $\mathbf{v}_{\boldsymbol{k}} = \partial E_{\boldsymbol{k}}/
\partial{\mathbf{k}}$; a proportionality factor
between $v_{\boldsymbol{k}}$ and $\mathbf{v}_{\boldsymbol{k}}^z$
related to a particular choice of
normalization of the Bloch waves is introduced further in Sec.IV.B.

  The potential $\Delta V(\mathbf{r})$ describing the constriction introduces
a perturbation to the perfect conductor. Let $\Phi_{\boldsymbol{k}}({\bf r},E)$
be a perturbed state which is a solution of the Lippmann-Schwinger equation for
an in-coming state in~L:
\begin{eqnarray}
\label{LS}
\Phi_{\boldsymbol{k}}({\bf r},E) & = &
{\Phi}^{\circ}_{\boldsymbol{k}}({\bf r},E) \\
\nonumber
& & + \int
d^3\mathbf{r'}\ G^{+}_0(\mathbf{r},\mathbf{r'},E) \Delta
V(\mathbf{r'}) \Phi_{\boldsymbol{k}}(\mathbf{r'},E)
\end{eqnarray}
where the integral goes over all space, and
$G^{+}_0(\mathbf{r},\mathbf{r'},E)$ is the retarded
Green's function of the perfect conductor.
Asymptotic behavior of $\Phi_{\boldsymbol{k}}({\bf r},E)$ is
\begin{eqnarray}
\nonumber
\Phi_{\boldsymbol{k}}({\bf r},E){\Bigl. \Bigr|}_{z\to -\infty}
& = & \Phi^{\circ}_{\boldsymbol{k}}({\bf r},E) +
\sum_{\boldsymbol{k'}}\rho_{\boldsymbol{k k'}}(E)\
{\Phi}^{\circ\,*}_{\boldsymbol{k'}}({\bf r},E) \\
\label{phi_k_limit}
\Phi_{\boldsymbol{k}}({\bf r},E){\Bigl. \Bigr|}_{z\to +\infty}
& = & \sum_{\boldsymbol{k'}} \tau_{\boldsymbol{k k'}}(E)\
{\Phi}^{\circ}_{\boldsymbol{\kappa'}}({\bf r},E)
\end{eqnarray}
where $\tau_{\boldsymbol{kk'}}(E)$ and
$\rho_{\boldsymbol{kk'}}(E)$ are transmission and
reflection amplitudes, assuming elastic scattering
($E_{\boldsymbol{k}} = E = E_{\boldsymbol{k}'}$).
According to the Landauer-B\"uttiker formula,\cite{Landauer} conductance
is given by $g = g_0\mathrm{Tr}(\tau\tau^{\dagger})$, where
trace goes over in-coming states ($\boldsymbol{k}$) in the left electrode and $E=E_F$.
An equivalent formulation with respect to in-coming states ($-\boldsymbol{k'}$)
from the right electrode reads as
$g = g_0\mathrm{Tr}(\tau^{\dagger}\tau)$.

Eigenchannels appear from a unitary transformation of in- and
out-states. Let $\omega$ be a unitary transform of in-states in~L:
$
%\label{in_nu}
\Phi^{\circ\,\mathrm{in}}_{\nu}({\bf r},E)
|_{z\to -\infty} = \sum_{\boldsymbol{k}}
\omega_{\nu\boldsymbol{k}}(E)\,
\Phi^{\circ\,\mathrm{in}}_{\boldsymbol{k}}({\bf r},E)$.
The corresponding solution $\Phi_{\nu}({\bf r})$
of Eq.~(\ref{LS}) for an arbitrary {\bf r}~is
\begin{equation}
\label{phi_chann}
\Phi_{\nu}({\bf r},E) = \sum_{\boldsymbol{k}}
\omega_{\nu\boldsymbol{k}}(E)\,\Phi_{\boldsymbol{k}}({\bf r},E),
\end{equation}
The unitary transform $\omega$ is defined such way that
the transmission matrix $T = \tau\tau^{\dagger}$ is
diagonal in the basis~$\nu$:
\begin{equation}
\label{omega}
\omega\, (\tau \tau^{\dagger})\,\omega^{\dagger} =
\mbox{diag}\{T_{\nu}\},
\end{equation}
and the conductance reads as $g = g_0 \sum_{\nu} T_{\nu}(E_F)$,
where the $T_{\nu}$'s are transmission probabilities
of eigenchannels.

The matrix $\tau$, however, is not diagonal in basis $\nu$.
Following Ref.~\onlinecite{Brandbyge1} one
can introduce a unitary matrix $\theta$ which satisfies the equation:
\[
 \omega\tau \theta^{\dagger}  =
 \theta \tau^{\dagger}\omega^{\dagger} =
 \mbox{diag}\{\sqrt{T_{\nu}}\},
\]
where all quantities are energy dependent. The solution is
$\theta = \mbox{diag}\{1/\sqrt{T_{\nu}}\} \omega \tau $.
The following properties of $\theta$ can be checked:
$\theta\theta^{\dagger} = \delta_{\nu\nu'}$,\
$\theta^{\dagger}\theta = \delta_{\boldsymbol{kk'}}$,
thus $\theta$ is indeed the unitary matrix.
It diagonalizes $\tau^{\dagger}\tau$:
\begin{equation}
\label{theta}
\theta\,(\tau^{\dagger}\tau)\, \theta^{\dagger}
= \mathrm{diag}\{T_{\nu}\}.
\end{equation}
Matrix $\theta$ performs a unitary transform of out-states in~R,
so that the linear combination $\nu$ of in-coming states in~L
scatters into the linear combination $\nu$ of the out-states in~R,
$
\Phi^{\circ\,\mathrm{out}}_{\nu}(\mathbf{r},E)|_{z \to +\infty} =
\sum_{\boldsymbol{k}} \theta_{\nu\boldsymbol{k}}(E)\,
\Phi^{\circ\,\mathrm{out}}_{\boldsymbol{k}}({\bf r},E),
$
with the transmission amplitude $\sqrt{T_{\nu}(E)}$, namely:
$\Phi_{\nu}(\mathbf{r},E)|_{z\to +\infty} =
\sqrt{T_{\nu}(E)}\ \Phi^{\circ\,\mathrm{out}}_{\nu}(\mathbf{r},E)$.

 One can show\cite{BarStone,Mavropoulos} that
for the Bloch states at the same energy ($E_{\boldsymbol{k}} = E
= E_{\boldsymbol{k'}}$) in the ideal leads
the following relations hold for the current matrix elements:
\begin{equation}
\label{Curr_Bloch1}
\displaystyle \int_S dS \left[
\Psi^{\circ}_{\boldsymbol{k}}(\mathbf{r},E)\,
i\!\stackrel{\leftrightarrow}%
{\partial_z}
\Psi^{\circ\,*}_{\boldsymbol{k'}}(\mathbf{r},E)
\right] \ =\
\frac{v_{\boldsymbol{k}}}{2\pi}
\, \delta_{\boldsymbol{kk'}},
\end{equation}
where the Bloch waves are either left ($v_{\boldsymbol{\kappa}}
\propto (\partial E_{\boldsymbol{k}}/ \partial\mathbf{k})_z >0$) or
right-travelling ($v_{\boldsymbol{\kappa}} < 0$),
the operator $\stackrel{\leftrightarrow}\partial_z$ is
defined as $f\stackrel{\leftrightarrow}\partial_zg =
f (\partial_z g) - (\partial_z f)g$,
and the integral goes over infinite plane $S$ (cross-section of the lead)
which is perpendicular to the current direction~$z$.

  In case of the perturbed system the orthogonality relation
holds for current matrix elements in the basis of eigenchannels.
Using Eqs.~(\ref{phi_k_limit}), (\ref{omega}) and (\ref{Curr_Bloch1})
we can compute it in the asymptotic region of the right (R)
lead:\cite{conjphi}
\begin{equation}
\label{Curr_Right}
\int\limits_{S_R} dS \left[
\Phi_{\nu}(\mathbf{r},E)\,
i\!\stackrel{\leftrightarrow}%
{\partial_z} {\Phi}^{*}_{\mu}(\mathbf{r},E)
\right]_{z\to +\infty} =
\frac{T_{\nu}(E)}{2\pi}\,\delta_{\nu\mu}.
\end{equation}
We note, that Eq.~(\ref{Curr_Right}) holds for {\it any} position $z$ of
the plane $S$. Because the wave functions of channels are solutions
of the Schr\"odinger equation with a real potential corresponding
to the same energy, the flux through arbitrary plane $S$ is conserved
(a proof is similar to that of Appendix~A of Ref.~\onlinecite{Mavropoulos}).

\section{Conduction eigenchannels within the KKR method}

\subsection{Evaluation of conductance}

To calculate the ballistic conductance, we employ the Kubo linear
response theory as formulated by Baranger and Stone: \cite{BarStone}
\begin{eqnarray}
\nonumber
g & =& g_0
%\frac{e^2\hbar^3}{8\pi m^2}
\int_{S_{\mathrm{L}}}dS\int_{S_{\mathrm{R}}}dS\,'
\\ & & \times\ G^{+}({\bf r},{\bf r'},E_F)
\stackrel{\leftrightarrow}\partial_z
\stackrel{\leftrightarrow}\partial_{z'} G^{-}({\bf r'},{\bf r},E_F),
\label{cond_g}
\end{eqnarray}
where $G^{-}$ and $G^{+}$ are retarded and advanced
Green's functions, respectively, and $g_0 = 2e^2/h$.
The integration is performed over left ($S_{\mathrm{L}}$) and right
($S_{\mathrm{R}}$) planes which connect
the leads with the scattering region (Fig.~1). The current flows in $z$ direction.
The implementation of Eq.~(\ref{cond_g}) within the KKR method, related convergence tests
and further details were discussed in recent publication.\cite{Mavropoulos}

 In this paper we present a further extension of Ref.~\onlinecite{Mavropoulos},
a method for the evaluation of conduction eigenchannels. For that,
we will follow closely the analysis of
Refs.~\onlinecite{BarStone,Mavropoulos} where the equivalence of
the Kubo and Landauer approaches to conductance problem was shown.
We proceed in four steps: (i) we remind the KKR representation of
the Bloch functions; (ii) we build up the asymptotic expansion of
the Green's function in terms of unperturbed Bloch states of the
leads; (iii) we construct the transmission matrix $T
=\tau\tau^{\dagger}$ in $\boldsymbol{k}$-space; (iv) finally, we
find an equivalent representation of the transmission matrix
within the local KKR basis. Further solution of the eigenvalue
problem leads us to conduction channels.  We mention here one
aspect of the problem: for realistic calculations the planes
$S_{\mathrm{L}}$ and $S_{\mathrm{R}}$ (see Fig.~2 for details) are
usually placed in finite distance from the atomic constriction.
Nevertheless, we first focus on the asymptotic limit before we
discuss the  realistic situation which is considered in Sec.~IV.E
and Appendix~A.

\subsection{Atomic orbitals and Bloch functions}
Let $\mathbf{r}$ be arbitrary point in the asymptotic
region of the lead (Fig.~2). In the KKR method the local,
energy dependent basis of atomic orbitals is defined at each unit cell $n$:
\begin{eqnarray}
\label{KKR_AO}
\phi^n_L(\mathbf{r},E) & = &
\phi_L(\mathbf{r}-\mathbf{R}_n,E)\\
\nonumber
& =  &  R_L(\mathbf{r}-\mathbf{R}_n,E)\, \Theta(\mathbf{r}-\mathbf{R}_n),
\end{eqnarray}
where $R_L$'s are real regular solutions \cite{RL_functions}
of the Schr{\"o}dinger equation for the potential $V_n(\mathbf{r})$
at cell $n$, and $\Theta$--function is 1 inside cell $n$ and is 0 outside it.
The unperturbed Bloch function
is given by expansion over atomic orbitals at all sites $n$ in the
Born-von K\'arm\'an supercell:
\begin{equation}
\label{psik_nL}
\Psi^{\circ}_{\boldsymbol{k}}(\mathbf{r},E) = \sum_{nL}
C_{\boldsymbol{k},nL}(E)\, \phi_L(\mathbf{r}-\mathbf{R}_n,E)
\end{equation}
with $C_{\boldsymbol{k},nL}(E) = e^{+i\mathbf{k}\mathbf{R}_n}
C_{\boldsymbol{k}L}^{\,\circ}(E)$.
Here the common notation $\boldsymbol{k} = (\mathbf{k},\lambda)$
for the Bloch vector $\mathbf{k}$ of the 1st Brillouin zone (BZ)
and the band index $\lambda$ is used.
The $C_{\boldsymbol{k}L}^{\,\circ}(E)$ are solutions of the
KKR band structure equations \cite{KKR_band_str} with energy $E$.

Considering a waveguide geometry, we will assume the
Bloch functions to be normalized per cross section of
the Born-von K\'arm\'an supercell (see Fig.~2)
with open boundary conditions along $z$-axis. Thus,
$C_{\boldsymbol{k},nL}(E) \propto 1/\sqrt{N_xN_y}$
with $N_xN_y$ being number of atoms per cross-section,
and orthogonality condition for Bloch waves
takes a form \cite{Mavropoulos}:
\begin{eqnarray*}
%\label{Psik_orthog}
\int\limits_V d^3\mathbf{r} \
\Psi^{\circ}_{\boldsymbol{k}}(\mathbf{r},E)
\Psi^{\circ\,*}_{\boldsymbol{k'}}(\mathbf{r},E')
& = & \frac{V}{(N_xN_y)} \delta_{\boldsymbol{k}\boldsymbol{k'}} \\
& = & 2\pi A_0\, \delta(k_z - k'_z)\,
\delta_{\boldsymbol{k_{\parallel}k'_{\parallel}}} \\
& = & |{v}_{\boldsymbol{k}}|\,
\delta(E-E')\,\delta_{ss'}\,
\delta_{\boldsymbol{k_{\parallel}k'_{\parallel}}}.
\end{eqnarray*}
Here $V = L_x L_y L_z$ is a volume of the supercell,
$A_0 = L_x L_y/N_xN_z$ is an area of the $xy$-unit cell
in the electrode, $s,s' =  \pm  1$ are signs of $k_z$ and $k'_{z}$,
relation $L_z\, \delta_{k_z k'_z} = 2 \pi\, \delta(k_z - k'_z) $
has been used, and velocity $v_{\boldsymbol{k}}$ along current flow
is defined as $v_{\boldsymbol{k}} = 2\pi A_0
(\partial E_{\boldsymbol{k}}/ \partial\mathbf{k})_z$.

\begin{figure}[t]
\begin{center}
\includegraphics[scale = 0.35]{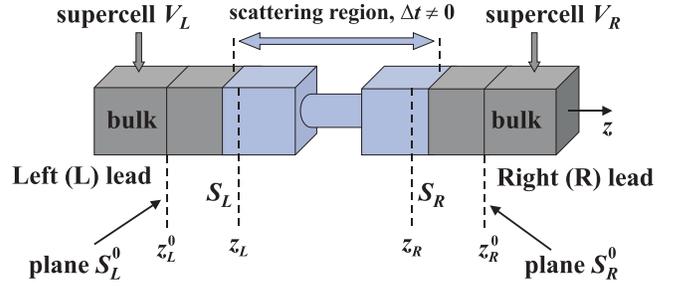}
\caption{Sketch of the system under consideration.
In a formal theory, left ($S^0_{\mathrm{L}}$) and right ($S^0_{\mathrm{R}}$) planes
are placed within the leads far away from a scattering region
(asymptotic limit). For all points $\mathbf{r}$ ($z < z^0_{\mathrm{L}}$)
and $\mathbf{r'}$ ($z' > z^0_{\mathrm{R}}$) within the Born-von K\'arm\'an
supercells (cubes) $V_{\mathrm{L}}$ and $V_{\mathrm{R}}$ asymptotic properties
are achieved. When implementation of the method is considered,
the conductance is calculated between the planes $S_{\mathrm{L}}$ and $S_{\mathrm{R}}$
positioned somewhere in the scattering region. }
\end{center}
\end{figure}

 Since the Bloch waves form a complete set, a back transform of
Eq.~(\ref{psik_nL}) exists:
\begin{equation}
\label{phi_nL}
\phi_L(\mathbf{r}-\mathbf{R}_n,E) =
\sum_{\boldsymbol{k'}} B^{\dagger}_{nL,\boldsymbol{k'}}(E)\,
\Psi^{\circ}_{\boldsymbol{k'}}(\mathbf{r},E_{\boldsymbol{k'}}),
\end{equation}
where $\boldsymbol{k'}$ sum runs over all ${\mathbf{k'}}$-points
in the 1st BZ and over all bands $\lambda'$. The $\dagger$ symbol
means Hermitian conjugate.
The expression for $B_{nL,\boldsymbol{k}}^{\dagger}$
can be obtained \cite{B_matrix} from known matrix $C_{\boldsymbol{k},nL}$.
One can prove further, that $C_{\boldsymbol{k},nL}$
and $B_{nL,\boldsymbol{k}}^{\dagger}$
obey the following orthogonality relations:
\begin{eqnarray}
\nonumber
\sum_{nL} C_{\boldsymbol{k},nL}(E)\,
B_{nL,\boldsymbol{k'}}^{\dagger}(E)
 & = & \delta_{\boldsymbol{kk'}}, \\
\label{BC}
\sum_{\boldsymbol{k}} B_{nL,\boldsymbol{k}}^{\dagger}(E)\,
C_{\boldsymbol{k},n'L'}(E) & = & \delta_{nn'}\delta_{LL'},
\end{eqnarray}
where in the second equation a sum over $\boldsymbol{k}$
is restricted to states with $E_{\boldsymbol{k}} =  E$.

\subsection{Asymptotic expansion of the Green's function}

Starting from the site angular momentum representation (\ref{Grr})
of the retarded Green's function within the KKR method
and using Eqs.~(\ref{psik_nL}) and (\ref{phi_nL})
we obtain the asymptotic expansion for $G^+(\mathbf{r},\mathbf{r'},E)$
over the unperturbed Bloch waves: \cite{BarStone,Mavropoulos}
\begin{eqnarray}
\label{G_knL}
\lefteqn{\left. G^+(\mathbf{r},\mathbf{r'},E)
\right|_{z,z' \to \mp \infty}  = } \\
\nonumber
& &\qquad\qquad \sum_{\boldsymbol{kk'}}
\Psi^{\circ\,*}_{\boldsymbol{k}}(\mathbf{r},E)\,
{\mathscr{A}}_{\boldsymbol{kk'}}(E)\,
\Psi^{\circ}_{\boldsymbol{k'}}(\mathbf{r'},E),
\end{eqnarray}
with $\mathbf{r}\in V_{\mathrm{L}}$, $\mathbf{r'}\in V_{\mathrm{R}}$
(see Fig.~2), and
\begin{eqnarray}
\label{A_kk}
\lefteqn{ {\mathscr{A}}_{\boldsymbol{kk'}}(E) = } \\
\nonumber
& & \sum_{n\in V_{\mathrm{L}}}\sum_{n'\in V_{\mathrm{R}}}\sum_{LL'}
B_{\boldsymbol{k},nL}(E)\,
G^{nn'}_{LL'}(E)\, B^{\dagger}_{n'L',\boldsymbol{k'}}(E)
\end{eqnarray}
or in a matrix form: ${\mathscr{A}} = B G B^{\dagger}$.
Formally, the $\boldsymbol{k}$-sums in Eq.~(\ref{G_knL}) are
performed over all $\mathbf{k}$-states in the 1st BZ and
over all bands $\lambda$. However, since the Green's function for
$\mathbf{r} \ne \mathbf{r'}$ is a solution of the Schr\"odinger
equation without a source term, only states
$\boldsymbol{k,k'}$ at the isoenergetic surface of energy $E$
contribute to the sum in the asymptotic
expansion.\cite{BarStone,Mavropoulos} Therefore,
\begin{equation}
\label{A0_kk}
{\mathscr{A}}_{\boldsymbol{kk'}}(E) =
\frac{1}{N_z^2}\
\overline{\delta}(E-E_{\boldsymbol{k}})\,
\overline{\delta}(E-E_{\boldsymbol{k'}})
A_{\boldsymbol{kk'}}(E),
\end{equation}
where
$\overline{\delta}(E-E_{\boldsymbol{k}}) =
\left({\Omega_0}/{S_{\lambda}}\right)\,
|\mathbf{v}_{\boldsymbol{k}}|\,
\delta(E-E_{\boldsymbol{k}})$.
Here $N_z$ is number of atom sites in Born-von K\'arm\'an supercell
along $z$ axis, $\Omega_0 = (2\pi)^3/V_0$ is volume of the 1st BZ,
$S_{\lambda}$ is the area of the isoenergetic
surface corresponding to band~$\lambda$,
$\mathbf{v}_{\boldsymbol{k}} =
\partial E_{\boldsymbol{k}}/\partial{\mathbf{k}}$.
For the discrete $\boldsymbol{k}$-points, the
function $(1/N_z)\overline\delta(E-E_{\boldsymbol{k}})$
equals 1 if $E = E_{\boldsymbol{k}}$,
and is 0 otherwise. In addition, boundary conditions
for the Green's function\cite{BarStone} constrain
matrix elements $A_{\boldsymbol{kk'}}(E)$
to be non-zeros only if $\boldsymbol{k}$
and $\boldsymbol{k'}$-states are right-travelling waves
(with positive velocity along $z$-axis) that corresponds
to the in-states $\boldsymbol{k}$
in the left lead and out-states $\boldsymbol{k'}$ in the
right one.

\subsection{Transmission matrix: asymptotic limit}

We proceed further, and use the asymptotic representation (\ref{G_knL})
of the Green's function to evaluate conductance according to Eq.~(\ref{cond_g}).
Assuming the integration planes to be placed within the leads infinitely
far from the scattering region, we obtain:
\begin{equation}
\label{g_tr_vAvA}
g  =  g_0 \mathrm{Tr}_{(\boldsymbol{k})}
\left[ V_{\mathrm{L}}\, {\mathscr{A}}(E_F)\, V_{\mathrm{R}}\,
{\mathscr{A}^{\dagger}}(E_F) \right],
\end{equation}
where the diagonal operators of velocities $V_{\mathrm{L}}$
and $V_{\mathrm{R}}$ (related to the left and right planes) acting
in the $\boldsymbol{k}$-space were introduced:
$
%\label{VV_k}
\bigl[V_{\mathrm{L(R)}}  \bigr]_{\boldsymbol{kk'}} =
\left({v_{\boldsymbol{k}}}/{2\pi}\right)
\delta_{\boldsymbol{kk'}}$.
Formally,  the trace (Tr) in Eq.~(\ref{g_tr_vAvA}) goes over
all $\boldsymbol{k}$-states and the Fermi surface is taken into account
by means of Eq.~(\ref{A0_kk}) where $E = E_F$.

The velocity operators can be decomposed
into sum of two operators related to the Bloch states
with positive and negative velocities along~$z$:
$
\hat V =  \hat V^{+} + \hat V^{-},
$
where $\hat V^{+}$ is nonzero for right-travelling waves only,
while $\hat V^{-}$ is nonzero for left-travelling ones.
In the asymptotic limit, only in-coming and out-going
$\boldsymbol{k}$-states with positive velocities contribute
to the sums in Eq.(\ref{g_tr_vAvA}).
Using the relation between expansion coefficients $A_{\boldsymbol{kk'}}$  and
the transmission amplitudes $\tau_{\boldsymbol{kk'}}$ derived in
Refs.~\onlinecite{BarStone,Mavropoulos},
\begin{equation}
\label{Akk}
A_{\boldsymbol{kk'}}(E)\ = \ -2\pi i\,
\frac{\tau_{\boldsymbol{kk'}}(E)}
{\sqrt{v_{\boldsymbol{k}} v_{\boldsymbol{k'}}}},
\end{equation}
we obtain:
\begin{eqnarray}
\label{g_vp_A_vp_A}
g & = & g_0 \mathrm{Tr\,}_{(\boldsymbol{k})}
\left[ V^{+}_{\mathrm{L}}\, {\mathscr{A}}(E_F)\,
V^{+}_{\mathrm{R}}\, {\mathscr{A}}^{\dagger}(E_F)\right] \\
\nonumber
{} & = & g_0 \mathrm{Tr}_{(\boldsymbol{k})}\,[T(E_F)],
\end{eqnarray}
where a representation of $T = \tau\tau^{\dagger}$
in $\boldsymbol{k}$-space is given~by
\begin{equation}
\label{TkE}
T(E) = (V^{+}_{\mathrm{L}})^{1/2}\, {\mathscr{A}}(E)\,
V^{+}_{\mathrm{R}}\, {\mathscr{A}}^{\dagger}(E)\,
(V^{+}_{\mathrm{L}})^{1/2}
\end{equation}
with a positive definite operator under square-roots.

The $\boldsymbol{k}$-representation is formal but not suitable for
implementation. To solve the eigenvalue problem for $T = \tau\tau^{\dagger}$
the mapping on the site-angular momentum $(n,L)$-space
of the KKR method should be presented. Such mapping is realized through the
expansion (\ref{psik_nL}) of the Bloch functions over atomic orbitals,
so that velocity operators in $\boldsymbol{k}$-space take a form:
\[
\bigl[V_{\mathrm{L(R)}}\bigr]_{\boldsymbol{kk'}} =
\sum_{nn'\in S^0_{\mathrm{L}}} \sum_{LL'}
C_{\boldsymbol{k},nL}\,
\bigl[D_{\mathrm{L(R)}}\bigr]^{nn'}_{LL'}\,
C^{\dagger}_{n'L',\boldsymbol{k'}},
\]
where site-diagonal operators $D_{\mathrm{R}}$ and $D_{\mathrm{L}}$
defined on atomic orbitals are the KKR-analogue
of velocities: \cite{KKR_velocity}
\begin{eqnarray}
\label{D_LR}
\lefteqn{\bigl[D_{\mathrm{L(R)}}(E)\bigr]^{nn'}_{LL'}\ =} \\
\nonumber\displaystyle
& & \pm\, \delta_{nn'}\, \int_{S^n_{\mathrm{L(R)}}} dS \left [R^n_L({\bf r},E)\,
i\stackrel{\leftrightarrow}{\partial_z}
R^{n}_{L'}({\bf r},E) \right]
\end{eqnarray}
here integral is restricted to the cross-section of the unit cell
around site~$n$. Now we can evaluate conductance according to
Eq.(\ref{g_tr_vAvA}). Taking into account that
$B^{\dagger}C = C^{\dagger} B = \delta_{nn'} \delta_{LL'}$
[Eq.(\ref{BC})], we obtain:
\begin{equation}
\label{g_TraceDGDG}
g = g_0 \mathrm{Tr}_{(n,L)}
\left(D_{\mathrm{L}}\, G\, D_{\mathrm{R}}\, G^{\dagger}
\right),
\end{equation}
here (and further) all matrices are assumed to be taken at the Fermi energy,
$G=\{G_{LL'}^{nn'}\}$ stands for matrix notation of the structural
Green's function introduced in Eq.~(\ref{Grr}),
and trace (Tr) involves sites and orbitals related to the atomic
plane~$S^0_{\mathrm{L}}$ (Fig.~2).

The operators  $D_{\mathrm{L}}$, $D_{\mathrm{R}}$ are anti-symmetric
and hermitian, thus their spectrum consists of pairs
of positive and negative eigenvalues: $\pm\delta^0_i$. Let $U$ be the unitary
transform of matrix $D$ (either $D_{\mathrm{L}}$ or $D_{\mathrm{R}}$) to
a diagonal form:
\begin{equation}
\label{D_to_D0}
D^0 = U^{\dagger} D U  = \mathrm{diag}\{\pm \delta^0_i\}.
\end{equation}
Decomposition of operator ${D}$ into two terms,
$D = D^{+} + D^{-}$, related to the right- and left-travelling waves
is naturally given in the basis of eigenvectors:
\begin{equation}
\label{D_UD0U}
D^{\pm} = U D^{0(\pm)} U^{\dagger}
\end{equation}
with
\[
D^{0(+)}\  = \ \left(
\begin{array}{cc}
\Delta^{+} & 0 \\
0 & 0
\end{array}
\right), \qquad
D^{0(-)}\  = \ \left(
\begin{array}{cc}
0 & 0 \\
0 & -\Delta^{-}
\end{array}
\right),
\]
where $\Delta^{+}$, $\Delta^{-}$ are positive (non-negative)
diagonal matrices. The analogue of Eq.~(\ref{g_vp_A_vp_A}) reads as
\begin{equation}
\label{g_Dp_G_Dp_G}
g|_{z,z' \to \mp \infty} = g_0 \mathrm{Tr}_{(n,L)}
\left(
D^{+}_{\mathrm{L}}\, G\, D^{+}_{\mathrm{R}}\, G^{\dagger}
\right).
\end{equation}

Now we are ready to build the $(n,L)$-representation
for transmission matrix $T = \tau\tau^{\dagger}$.
For that, one should extract a square-root from the positive
definite operator
$
V^{+}_{\mathrm{L}} =
C D^{+}_{\mathrm{L}} C^{\dagger}
$
defined on $\boldsymbol{k}$-space with help
of {$(n,L)$}-space. Namely, because the $V^{+}_L$ is positive definite,
it can be represented in the following form: \cite{Gantmacher}
$ V^{+}_{\mathrm{L}} = \Omega_{\mathrm{L}} \Omega^{\dagger}_{\mathrm{L}}$.
Here an operator $\Omega_{\mathrm{L}}$
maps the $\boldsymbol{k}$-space on the {$(n,L)$}-space,
$
 \Omega_{\mathrm{L}}: \{\boldsymbol{k}\} \to \{n,L\}$.
The solution for $\Omega_L$ is
\[
\Omega_{\mathrm{L}} = C (D^{+}_{\mathrm{L}})^{1/2}\,\cal{E},
\]
where $\cal{E}$ is an arbitrary unitary matrix
(${\cal{E}\cal{E}^{\dagger}}=1$) and
$ (D^{+}_{\mathrm{L}})^{1/2} =
U_L [ D^{0(+)}_{\mathrm{L}} ]^{1/2} U^{\dagger}_L
$
with $[D^{0(+)}_{\mathrm{L}}]^{1/2}$ being square-root of
the positive definite diagonal matrix [Eq.(\ref{D_UD0U})].

To find $T = \tau\tau^{\dagger}$, we start from
Eq.~(\ref{g_vp_A_vp_A}) and proceed as follows:
\begin{eqnarray*}
g & = & g_0 \mathrm{Tr\,}_{(n,L)} \left(
\Omega_{\mathrm{L}} \Omega^{\dagger}_{\mathrm{L}} \,
{\cal{A}}\, V^{+}_{\mathrm{R}} {\cal{A}}^{\dagger}
\right) \\
 & = &  g_0 \mathrm{Tr\,}_{(n,L)} \left(
\Omega^{\dagger}_{\mathrm{L}} \, {\cal{A}}\,
V^{+}_{\mathrm{R}} {\cal{A}}^{\dagger}
\Omega_{\mathrm{L}} \right) \\
& = & g_0 \mathrm{Tr\,}_{(n,L)}
\left[ (D^{+}_{\mathrm{L}})^{1/2}C^{\dagger}
\bigl(B G {B}^{\,\dagger} \bigr)
\bigl({C} D^{+}_{\mathrm{R}}{C}^{\dagger} \bigr)
 \right. \\
& & \left. \times \bigl({B} G^{\,\dagger} B^{\dagger} \bigr)
C (D^{+}_{\mathrm{L}})^{1/2}\,
\right]
\end{eqnarray*}
where Eq.(\ref{A_kk}) was used. Because of
$B^{\dagger} C = \delta_{nn'}\delta_{LL'}$
[Eq.~(\ref{BC})], we obtain:
\begin{equation}
\label{g_Tr_nL}
g =  g_0 \mathrm{Tr\,}_{(n,L)} ({\cal{T}} ),
\end{equation}
where, in the asymptotic limit,
the transmission matrix $T = \tau\tau^{\dagger}$ in
$(n,L)$-representation is given by
\begin{equation}
\label{T_nL_limit}
{\cal{T}} = (D^{+}_{\mathrm{L}})^{1/2}
G\, D^{+}_{\mathrm{R}}\, G^{\,\dagger} (D^{+}_{\mathrm{L}})^{1/2}.
\end{equation}
The trace in Eq.(\ref{g_Tr_nL}) goes over all sites $n$ and
orbitals~$L$ of the atomic plane $S^0_{\mathrm{L}}$ in the left
lead (Fig.~2). Equivalent formula can be written for the right
lead. To conclude, one can prove that spectrum of the obtained
matrix coincides with spectrum of transmission matrix $T$
[Eq.~(\ref{TkE})] defined in $\boldsymbol{k}$-space (see
Appendix~B for details). Therefore, solution of the eigenvalue
problem for $\cal{T}$ gives us required transmission
probabilities.

\begin{figure}[t]
\begin{center}
\includegraphics[scale = 0.5]{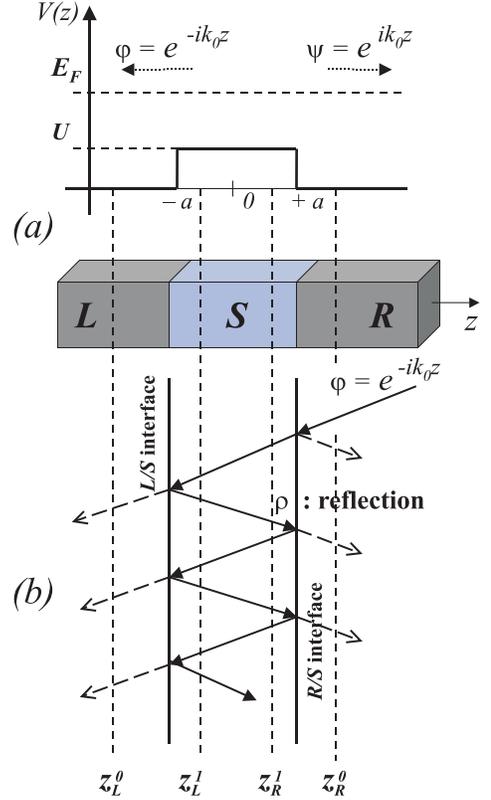}
\caption{Top (a): A model step-like potential for
the free-electron gas within the $L/S/R$ structure
having two dimensional periodicity. The electrons moving with energy $E = E_F$
from the leads (L and R) are scattered at the potential step~$U$.
The $\boldsymbol{k_\parallel}$ momentum is conserved, while the momentum $k_z$ along
$z$-axis is $k_0 = \left[2mE - \boldsymbol{k}_{\parallel}^2 \right]^{1/2}$
in the leads, and $k_1 = \left[2m(E-U) - \boldsymbol{k}_{\parallel}^2\right]^{1/2}$
in the S layer. Bottom (b): A multiple reflection of the in-coming
from the R lead wave $e^{-i k_0 z}$ within
the spacing layer $S$, where it is a linear combination
$ \phi(z) = \chi_{+}(z) + \chi_{-}(z) = \alpha\, e^{-i k_1 z} + \beta\, e^{+i k_1 z}$
of two functions. If $z_L^1$ is position of one of the planes taken for the conductance
evaluation, transmission $T(\boldsymbol{k}_{\parallel})$ of channel
$\boldsymbol{k}_{\parallel}$ is proportional to the current
$j = \phi(z)\ i\!\stackrel{\leftrightarrow}{\partial_{z}} \phi^{*}(z) =
2k_0 = 2k_1 |\alpha|^2 - 2k_1|\beta|^2 = j_{+} - j_{-} $,
which is sum of two terms (positive and negative)
due to contributions from two functions,
$\chi_{+}(z)$ and $\chi_{-}(z)$. The value
$\rho = \beta/\alpha = (k_1-k_0)/(k_1 + k_0) e^{+2ik_1 a}$
has a meaning of the reflection coefficient of the S/R or S/L interface.
However, if plane is chosen in the asymptotic region of the lead,
at point $z_{\mathrm{L}}^0$, transmission $T(\boldsymbol{k}_{\parallel}) \sim 2k_0$
contains only one positive contribution.}
\end{center}
\end{figure}

\subsection{General case: arbitrary positions of planes}

In practical calculations of conductance with the use of the Kubo formula,
the left (L) and right (R) planes are positioned {\it somewhere}
in the leads (Fig.~2). Expression (\ref{g_TraceDGDG}) is valid in general
case and result is exactly the same as in
Ref.~\onlinecite{Mavropoulos}. Operator $D_{\mathrm{L}}$
in Eq.(\ref{g_TraceDGDG}) is sum of two terms:
$D_{\mathrm{L}} =  D^{+}_{\mathrm{L}} + D^{-}_{\mathrm{L}}$.
Therefore, we can write down
\begin{eqnarray}
\nonumber
g & = & g_0 \mathrm{Tr}
\left(D^{+}_{\mathrm{L}}\, G\, D_{\mathrm{R}}\, G^{\dagger} \right)
+ g_0 \mathrm{Tr} \left(D^{-}_{\mathrm{L}}\, G\, D_{\mathrm{R}}\, G^{\dagger}
\right) \\
\label{2_terms}
& = & g^{+} + g^{-},
\end{eqnarray}
where $g^{+}$ and $g^{-}$ denote two contributions. In a formal
theory, when the atomic plane $S_{\mathrm{L}}$ is placed in the
asymptotic region of the left lead far from the atomic
constriction the second term $g^{-}$ in Eq.(\ref{2_terms}) is
equal to zero. In practice, the real space summation of current
contributions includes only a finite number of sites at both
atomic planes, because the current flow along $z$ direction is
localized in the vicinity of the contact. Due to numerical effort
we are forced to take integration planes closer to the
constriction in order to obtain convergent value for the
conductance with respect to number of atoms included in summation.
In addition, even better convergence for matrix elements is
required to solve the eigenvalue problem. A compromise can be
usually achieved but positions of the atomic planes
$S_{\mathrm{L}}$ and $S_{\mathrm{R}}$ do not meet the asymptotic
limit criterion. However, since the electron current through the
structure is conserved, any position of the planes is suitable for
the calculation of conductance. If $S_{\mathrm{L}}$ is placed
somewhere in the scattering region we have to sum up all multiple
scattering contributions. We show in Appendix~A, that all multiple
scattering contributions in direction of the current cause
$g^{+}$, whereas all scattering contributions in opposite
direction give rise to $g^{-}$. Thus, the first term, $g^{+}$, in
Eq.~(\ref{2_terms}) is always positive, while the second one,
$g^{-}$, is always negative. To make this statement clear, an
illustration of scattering events is shown in Fig.~3 assuming a
simple free-electron model. In the region of the lead where the
potential is a small perturbation with respect to the bulk
potential the contribution to the conductance due to $g^{-}$ is
one order of magnitude smaller than $g^{+}$.

  To find transmission probabilities of eigenchannels,
one has to apply the procedure introduced in the previous section
independently for both terms, $g^{+}$ and $g^{-}$. We refer to
Appendix~A for a mathematical justification. Expression for
conductance takes a form:
\begin{equation}
\label{g_2terms}
g =  g_0 \mathrm{Tr\,}_{(n,L)} ({\cal{T}}^{+})
  +  g_0 \mathrm{Tr\,}_{(n,L)} ({\cal{T}}^{-}),
\end{equation}
with
\begin{equation}
\label{TT}
{\cal{T}}^{\pm} = \pm {(\pm D^{\pm}_{\mathrm{L}})}^{1/2}\,
G\, D_{\mathrm{R}}\, G^{\,\dagger} {(\pm D^{\pm}_{\mathrm{L}})}^{1/2}.
\end{equation}
We show in Appendix A that all eigenvalues of ${\cal T}^{+}$ are
either positive or zeros whereas all eigenvalues of ${\cal T}^{-}$
are negative or zeros.  To identify transmission probabilities
$T_n$ of channels the spectra of operators ${\cal T}^{+}$ and
${\cal T}^{-}$ have to be arranged in a proper way.  Then
transmission of the $n$-th channel is given by $T_n = \tau^{+}_n -
\tau^{-}_n$, where $\pm \tau_n^{\pm}$ are positive and negative
eigenvalues of the operators ${\cal T}^{+}$ and ${\cal T}^{-}$,
respectively. The $T_n$ does not depend on the positions of the
left ($S_{\mathrm{L}}$) and right ($S_{\mathrm{R}}$) planes, while
$\tau^{\pm}_n$ are $z$-dependent. In the asymptotic limit
$\tau^{-}_n|_{z_L\to -\infty} \to 0$ and $T_n =
\tau^{+}_n|_{z_L\to -\infty}$, so that the Landauer picture is
restored.

In general case the direct way to find the pairs of eigenvalues is not evident
without a back transform to the $\boldsymbol{k}$-space.  However,
from the point of view of applications to the extremely small symmetric
atomic contacts, as the ones we are studying in this work,
the problem is easy to handle.
Since the number of contributing eigenmodes is limited,
the pairs of eigenvalues can be found by symmetry
analysis of the eigenvectors of ${\cal T}^{+}$ and ${\cal T}^{-}$.
Namely, using the symmetry properties
of the structural Green's function $G^{nn'}_{LL'}$ and
current matrix elements  $D^{nn'}_{LL'}$ one can show
that channel's transmission $T_n$ bounded between 0 and 1 is defined
by eigenvalues $\tau_n^{+}$ and $\tau_n^{-}$ which belong
to the same irreducible representation of the symmetry point group.

\section{Applications of the method}

In recent papers \cite{Imp_Paper1,Imp_Paper2} we have verified
the method described here by studying systematic changes in the conductance
of metallic constrictions in the presence of defect atoms.
Illustrative examples presented below focus
on single-atom contacts made of pure metals such as Cu, Ni, Co and Pd.

Copper serves mainly for test purposes. It is a representative of the noble metals
and has electronic properties similar to Ag and Au, for which a lot of experimental
results \cite{Brandbyge,Krans,Exp_noble_metals,Ludoph,Enomoto,Yanson_PhD}
as well as DFT based calculations \cite{TB_models,Ab-initio_1} are available.
In particular, a large number of experiments
for alkali metals (Li, Na, K)~\cite{Ludoph,Yanson_PhD,Yanson_Nature}
and noble metals (Au, Ag, Cu),~\cite{Brandbyge,Exp_noble_metals,Ludoph,Enomoto,Yanson_PhD}
employing different techniques under room an liquid-He temperatures,
show that conductance histograms have a dominant peak very close
to one conductance quantum $G_0 = 2e^2/h$, and smaller
peaks close to integer values.

  However, for
transition metal contacts (examples of which are Ni, Co and Pd)
the situation differs significantly.\cite{Ludoph,Enomoto}
Only one broad maximum centered somewhere
between $1\,G_0$ and $3\, G_0$ is usually observed
in conductance histograms.\cite{Ruitenbeek,Yanson_PhD}
That is a signature of the nontrivial decomposition of
conductance consisting of more than one perfectly transmitting channel,
\cite{Cuevas_TB_model,Scheer_Nature} since for
transition metal atoms $d$ states of different symmetry
are available at the Fermi level. The question on half-integer conductance
quantization has been addressed.\cite{Half_int_FM,Half_int_exp} However,
recent experiments \cite{Untiedt} do not confirm this hypothesis
thus pointing to the conclusion that the electron transport through
ferromagnetic contacts can never be fully spin-polarized.
Another issue is a large magnetoresistance (MR) effect
\cite{Huge_MR,Huge_MR_1,Hua_Chopra} observed for metallic point contacts made of different
magnetic materials. This field is known to be full
of controversy. There is still a continuing discussion on whether
or not the enormous MR values found experimentally
could be of electronic origin
\cite{Electr_origin,Tagirov,Velev_Butler,PRB_MR,NiO_surface,Ni_Spain,NiO_Spain}
or the effect is just
due to atomic rearrangements in the neck region of a contact as a response
to the applied magnetic field.\cite{Magnetostriction1,Magnetostriction2}
Extensive discussion on this topic can be found in the recent
review paper by Marrows. \cite{Marrows} We will comment further on the
above issues.

\subsection{Computational details}

An atomic configuration of a constriction used in our calculations
is presented in Fig.~1. The single atom contact was modeled by a small cluster
attached to the semi-infinite fcc (001) leads.
The cluster consists of two pyramids joined via the vertex atoms separated
by a distance $a_0/\sqrt{2}$ . Here $a_0$ is the experimental lattice constant
of fcc metals: 6.83 a.u.\ for Cu, 6.66 a.u.\ for Ni, 6.70 a.u.\ for Co,
and 7.35 a.u.\ for Pd. The metals under consideration do not have a tendency
to form chains.\cite{Chains} An atomic bridge is most likely to be broken just
after the single atom limit is achieved. Thus, a configuration shown in Fig.~1
resembles one  of limiting configurations of point contacts which could appear
in the MCBJ experiments.

\begin{figure}[t]
\begin{center}
\includegraphics[scale = 0.75]{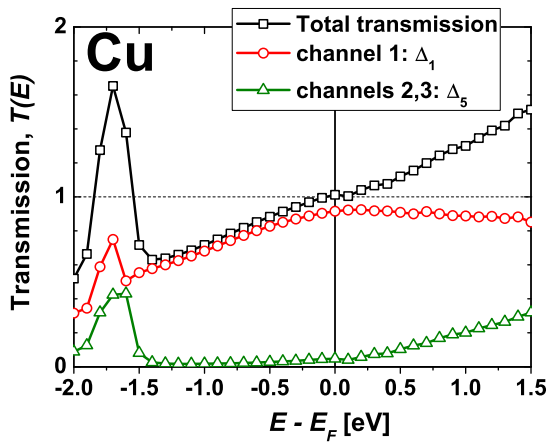}
\caption{(Color online) Energy-dependent transmission and its
decomposition to the conduction eigenchannels
for the Cu single atom contact shown in Fig.~1.}
\end{center}
\end{figure}

 Our calculations are based on DFT within
the local density approximation. The parametrization of Vosko, Wilk, and Nusair
\cite{Vosko_Wilk_Nusair} for the exchange and correlation
energy was used. The potentials were
assumed to be spherically symmetric around each atom
(atomic sphere approximation, ASA). However, the full charge
density, rather than its spherically symmetric part, was taken into account.
To achieve well converged results the angular momentum cut-off for the
wave functions and the Green's function was chosen to be $l_{\rm{max}}=3$ that
imposed a natural cut-off $2 l_{\rm{max}}=6$ for the
charge density expansion. In case of heavy element Pd the scalar
relativistic approximation\cite{SRA} was employed.
For the conductance calculation the surface Green's function was computed
using a small imaginary part ${\rm Im}E = 0.04\,{\rm mRy}$ and about
250.000 \mbox{k-points} were taken in the 2D Brillouin zone.
Instead of integration over planes,
current matrix elements (\ref{D_LR}) were averaged over atomic
layers as described in detail in Ref.~\onlinecite{Mavropoulos}.
A typical error in the calculation
of conductance was $ \sim 5\%$.

\subsection{Symmetry analysis of eigenchannels}

To understand the relation between the electronic structure and
the transport properties of atomic contacts
we consider the energy-dependent total transmission, $T(E)$,
and its decomposition to the conduction
eigenchannels, $T_i(E)$. Results are shown in Figs.~4,\,6,\,8,\,9 for
the case of Cu, Ni, Co and Pd point contacts, respectively.
The investigated structure (Fig.~1) has a $C_{4v}$ symmetry.
Further we denote individual channels by the indices
of irreducible representations of this group
using notations of Ref.~\onlinecite{Wigner}, common in band
theory. In addition, each channel can be classified
according to the angular momentum contributions
when the channel wave function is projected
on the contact atom of the constriction. This is very
helpful since the channel transmission can be related
to the states of the contact atom.\cite{Cuevas_TB_model}
For example, the identity representation $\Delta_1$
of the $C_{4v}$ group is compatible with the  $s$, $p_z$ and $d_{z^2}$ orbitals
(here the $z$ is perpendicular to the surface
and passes through the contact atom),
while the two-dimensional representation $\Delta_{5}$
is compatible with the $p_x$, $p_y$, $d_{xz}$,  $d_{yz}$ orbitals.
The basis functions of $\Delta_2$ and $\Delta_{2'}$ are $d_{x^2-y^2}$ and
$d_{xy}$ harmonics, respectively.

\subsection{Cu contacts}

The energy-dependent transmission of Cu atomic contact
(shown in Fig.~1) is presented in Fig.4 together
with the eigenchannel decomposition.
At the Fermi energy the calculated conductance value is $G = 1.01\ G_0$. It
mainly consists of one open channel
of $\Delta_1$ symmetry which arises locally from
$s$, $p_z$ and $d_{z^2}$ orbitals when the wave function is projected
on the contact atom. This result is in a good agreement with a lot
of experiments\cite{Ludoph,Yanson_PhD} mentioned previously as well as with other
calculations involving different approaches.\cite{TB_models,Ab-initio_1}
The additional twofold degenerate channel has $\Delta_5$ symmetry.
Transmission of this channel increases
at energies above the Fermi level ($E_F$)
together with an increase of the $p_x, p_y$
contribution to the local density of states (LDOS) at the contact atom.
However, at $E < -1.5$~eV below the $E_F$ the $\Delta_5$
channel is built mainly from the $d_{xz}, d_{yz}$ orbitals of the Cu atom.

We would like to point out that in case of noble metals
the conductance of single-atom contact
is not necessarily restricted to one channel.
An example of a configuration which has more
channels (but still has only one Cu atom at the central position)
is presented on the top of Fig.~5. Because
of the larger opening angle for incoming waves as
compared with the preceding case, conductance of such system
is $G = 2.57\,G_0$ with major contribution from four channels
(Fig.~5 and its caption).
The value $2.57\,G_0$ correlates with a position of the
third peak in the conductance histogram
of Cu, which is shifted from $3\,G_0$ to
smaller values.\cite{Yanson_PhD,Ludoph}
As it is seen from the presented example, conductance quantization
does not occur for the metallic atomic-sized contacts.
In general, even for noble metals,
conduction channels are only partially open\cite{Ludoph}
in contrary to the case of quantum point contacts realized
in the two-dimensional electron gas where a clear conductance
quantization was observed.\cite{2DEG}

\begin{figure}[t]
\begin{center}
\includegraphics[scale = 1.00]{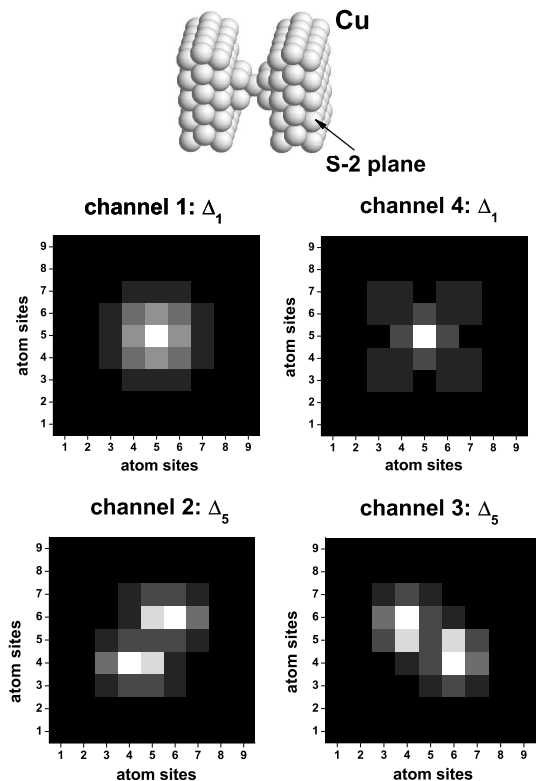}
\caption{ Wave functions $|\Psi_i(\mathbf{r})|^2$ (probability
densities) of the four dominating eigenchannels for the pyramidal
Cu contact shown on the top. Wave functions resolved to atoms are
visualized 2 atomic planes below the surface plane (S-2). Colors
from white to black correspond to consequently decreasing positive
values. Transmission probabilities of channels are: $T_1 = 0.90\
(\Delta_1)$,
$T_2 = T_3 = 0.71\ (\Delta_5)$, %
$T_4 = 0.08\ (\Delta_1)$, which are summed up to conductance $G =
2.57 G_0$. Further details are given in the text. }
\end{center}
\end{figure}

 For illustration, we present in Fig.~5 probability amplitudes of the eigenchannles
$|\Psi_i(\mathbf{r})|^2$ in real space resolved with respect to atoms
of the 2nd plane (S-2) below the surface. We see that the wave functions
of the 1st and the 4th channel with the highest symmetry
($\Delta_1$) obey all eight symmetry transformations of
the $C_{4v}$ group, while two different wave functions of
the double degenerate channel ($\Delta_5$) are transformed to each other
after some symmetry operations.

\begin{figure}[t]
\begin{center}
\includegraphics[scale = 0.75]{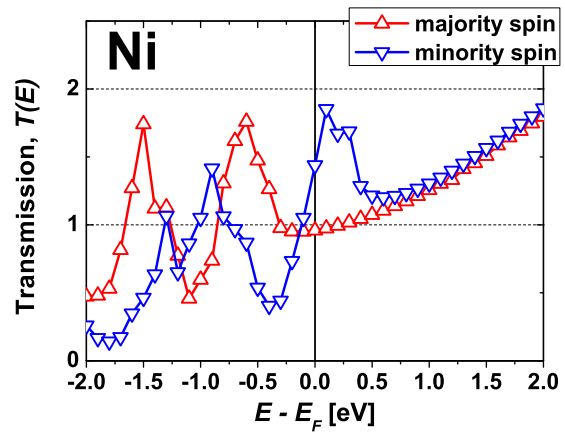}
\caption{(Color online) Spin-dependent transmission as a
function of energy for the Ni single atom contact shown in Fig.~1.}
\end{center}
\end{figure}

\subsection{Transition metal contacts}

We turn to transition metals, and consider
the ferromagnetic Ni assuming a uniform magnetization
of the sample. Transmission $T(E)$ split
per spin of a Ni contact is shown in Fig.~6.
A shift about $\sim 0.8$~eV along the energy axis between transmission curves is
seen that is in agreement with exchange-splitting of the Ni $d$ states.
Similar computational results regarding transmission of Ni constrictions
were reported by Solanki {\it et al}.,\cite{TB-LMTO-Ni}
Rocha {\it et al.}\cite{Rocha}\ and Smogunov {\it et al.}.\cite{Ni_Smogunov}
Exchange splitting estimated from their works
varies from 0.8~eV till 1.0~eV, but fine details differ
because of different atomic configurations and employed methodologies.
In this regard, exchange splitting about 2.0~eV observed
in transport calculations of Jacob {\it et al.}\cite{Ni_Spain} \ in case of Ni contact
seems to be overestimated.

The shift in energy due to different spins is observed as well
for the transmissions of individual channels (Fig.~7).
We see from Fig.~7 and Table~I that at the Fermi energy
the spin-up (majority) conductance of Ni contact is mainly determined by
one open $\Delta_1$ channel (similar to the case of Cu),
while three partially open channels, of $\Delta_1$ and $\Delta_5$ symmetry,
contribute to the spin-down (minority) conductance.
The minority $\Delta_5$ channel arises locally
from $d_{xz}$ and $d_{yz}$ states, rather than
from $p_x,p_y$ states whose contribution
to the spin-down LDOS at the Fermi energy is much smaller (Fig.~7).
The calculated conductance, $G = 1.20\,G_0$,
correlates with a position of the wide
peak in the conductance histogram of Ni centered
between $1\,G_0$ and $2\,G_0$.\cite{Yanson_PhD,Untiedt}

\begin{figure*}[b]
\begin{center}
\includegraphics[scale = 0.85]{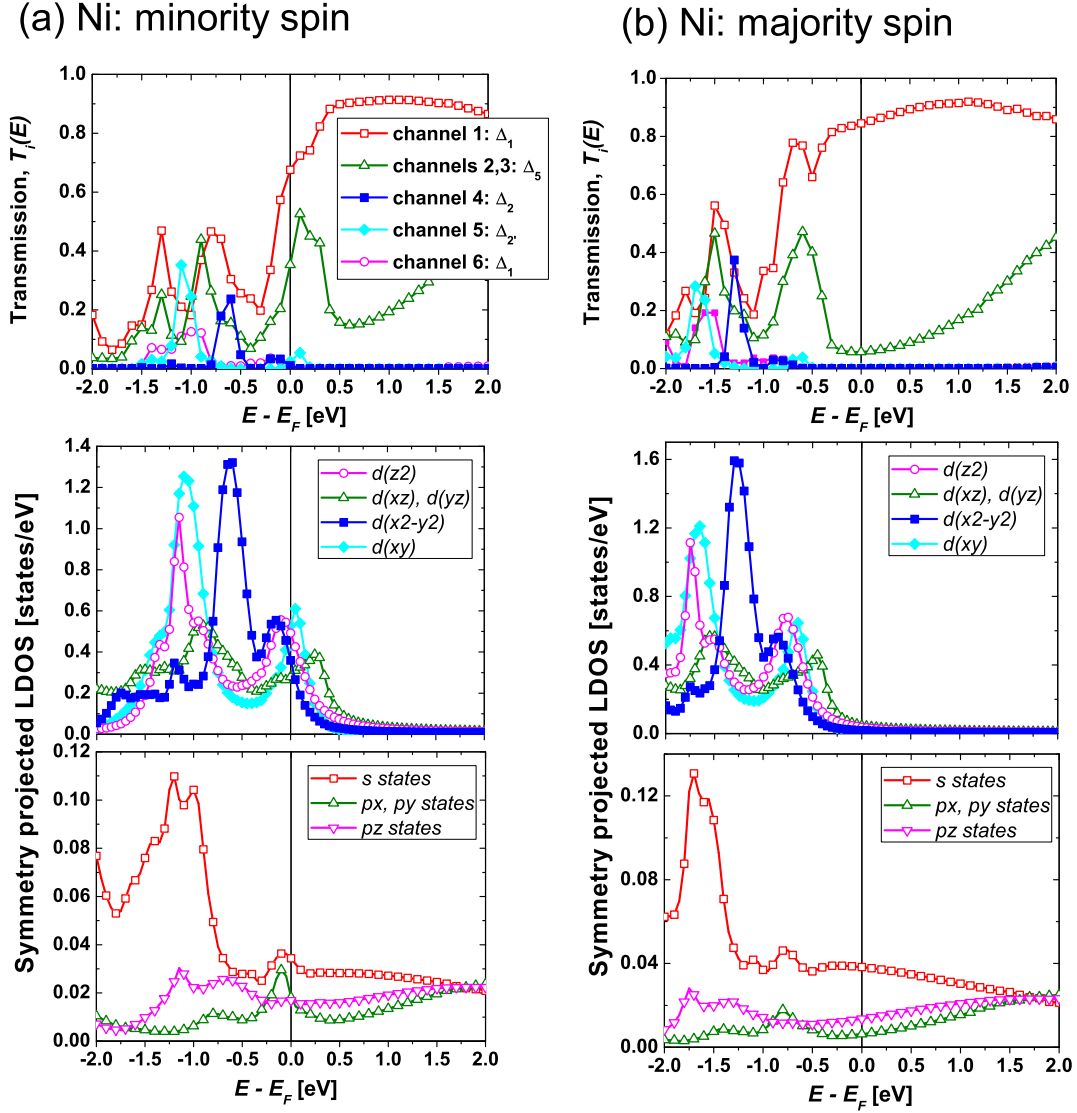}
\caption{(Color online)
Spin- and energy-dependent transmission decomposed to
conduction eigenchannels for the Ni atomic-sized constriction in comparison
with the symmetry projected local density of states at the contact atom
(i.e.\ apex atoms in Fig.~1).}
\end{center}
\end{figure*}

\begin{table*}[b]
\caption{Transmission probabilities of eigenchannels
at the Fermi energy of Ni and Co atomic
contacts shown in Fig.~1 for two different (P and AP)
orientations of magnetizations in the leads.
Only transmissions of the dominant channels are presented.
Magnetoresistance ratio defined as $\mathrm{MR} = (G_P - G_{AP})/G_{AP} \times 100\%$
is given in the last line.
}

\begin{ruledtabular}
\begin{tabular}{rcccccc}
{} & \multicolumn{3}{c}{Ni} & \multicolumn{3}{c}{Co} \\
\cline{2-4} \cline{5-7}
Channel & \multicolumn{2}{c}{P} & AP & \multicolumn{2}{c}{P} & AP \\
{} & ($\downarrow$)--spin & ($\uparrow$)--spin & ($\uparrow$) or ($\downarrow$) spin &
($\downarrow$)--spin & ($\uparrow$)--spin & ($\uparrow$) or ($\downarrow$) spin \\
\hline
$T_1\ (\Delta_1) $
   &  0.68  &  0.84  &  0.82  &  0.36  &  0.89  & 0.58 \\
$T_2 = T_3\ (\Delta_5)$
   &  0.35  &  0.06  &  0.31  &  0.14  &  0.07  & 0.09 \\
$T_4\  (\Delta_2)$
   &  {}   &  {}    &  {}     &  0.17  &   {}  & {} \\ \hline
{Transmission} &  1.44 & 0.96 & 1.45 & 0.83 & 1.03 & 0.76  \\ \hline
{MR ratio}  & {} & $-17\%$  & {} &  {} & $+23\%$   & {}  \\
\end{tabular}
\end{ruledtabular}
\end{table*}

Within the energy range shown in Figs.~6 and 7 ($\pm 2$~eV around
$E_F$), we count six eigenmodes of different symmetry
for both spins. At energies well above the Fermi level
($E > 1.0$~eV) the spin-splitting of $sp$ Ni states is lost,
and the picture is similar to what we have seen for Cu.
Three channels are present:
one open $\Delta_1$ ($sp_z$-like) channel with transmission around 0.9
and a partially open double degenerate $\Delta_5$ ($p_x,p_y)$
channel whose transmission increases monotonically as a function of energy.
However, below the Fermi energy all eigenmodes $T_i(E)$ display
a complicated behavior (Fig.~7, upper plots) that reflects
a complex structure of the LDOS projected on orbitals of
the contact atom (Fig.~7, bottom plots).
Below $E_F$ the existing $s$, $p_z$ and $d_{z^2}$ states
are strongly hybridized giving rise to two channels
of $\Delta_1$ symmetry at energies about $E \approx -1.0$~eV and
$E \approx -1.5$~eV for minority and majority spins, respectively.
A clear correlation between the symmetry projected LDOS and $T_i(E)$ is seen
for the pure $d$-channels, $\Delta_2$ ($d_{x^2-y^2}$) and $\Delta_{2'}$ ($d_{xy}$).
For example, the minority spin $d_{x^2 - y^2}$ resonance centered around $E = -0.7$~eV and
majority spin $d_{x^2 - y^2}$ resonance at $E = -1.3$~eV
reflect themselves as peaks in the transmission of the
minority and majority $\Delta_2$ channels.
The same is valid for the $d_{xy}$ states
at $E = -1.1$~eV (spin down) and $E = -1.7$~eV (spin up)
which cause the increase of transmission of the $\Delta_{2'}$
channel at the same energies. However, even at the resonances,
the $\Delta_2$ and $\Delta_{2'}$ channels are only
partially open because the $d_{x^2-y^2}$ and $d_{xy}$ orbitals
are spread perpendicular to wire ($z$-) axis
that prevents effective coupling with the neighboring atoms.

Our results for Co constriction are presented in Fig.~8.
As compared with Ni, the shift
between the spin-up and the spin-down $T(E)$ curves becomes larger
($\sim 1.7$~eV), because of the stronger exchange field of cobalt.
At the Fermi energy majority spin conductance is still dominated
by one highly transmitted $\Delta_1$ channel (Table~I), while for the
minority spin the $d_{x^2-y^2}$ resonance is pinned to the Fermi level
and results in the additional (as compared with Ni) channel
of $\Delta_2$ symmetry.
Thus four channels with moderate transmission probabilities
contribute to the minority spin conductance (Table~I).

\begin{figure}[t]
\begin{center}
\includegraphics[scale = 0.74]{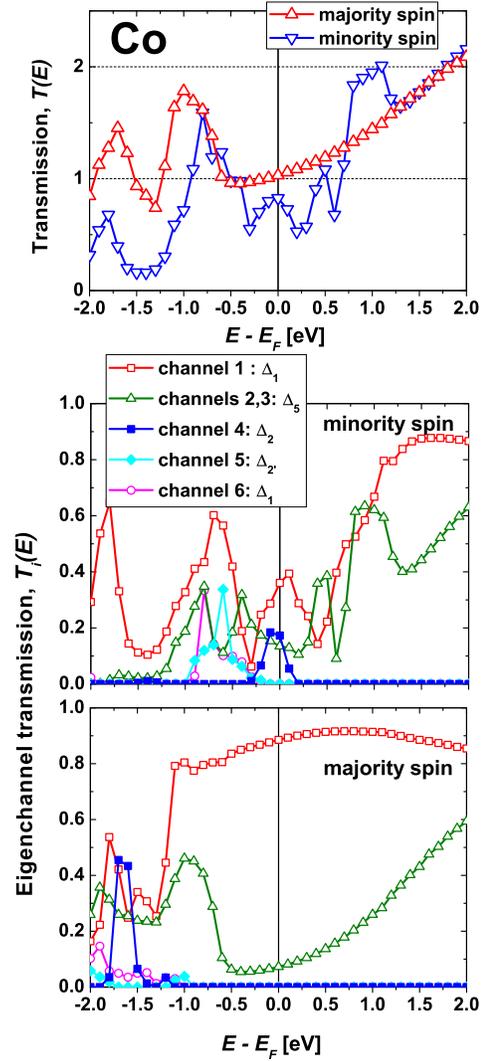}
\caption{(Color online) Spin- and energy-dependent transmission
decomposed to conduction eigenchannels for the Co single atom
contact shown in Fig.~1.}
\end{center}
\end{figure}

 Fig.~9 shows results for Pd.
According to recent theoretical predictions \cite{Pd_wires}
monatomic Pd wires might exhibit magnetic properties. However,
in this work we considered nonmagnetic solution,
since a coordination number even for
the contact atom was already
big enough (Fig.~1) to suppress magnetism.\cite{Pd_wires}
Pd is isovalent to Ni. The Fermi level crosses the partially filled $d$ band.
Therefore, the eigenchannel decomposition
resembles the minority spin channels of Ni.
However, due to a larger occupation number,
transmission curves are shifted $\sim 0.5$~eV downwards
in energy as compared with spin-down Ni modes.
The conductance $G = 1.41\,G_0$ is a sum of three channels.
This value is in agreement with the conductance histogram of Pd\cite{Pd_hist}
which shows a broad maximum around $ \simeq 1.7\, G_0$.

\begin{figure}[t]
\begin{center}
\includegraphics[scale = 0.75]{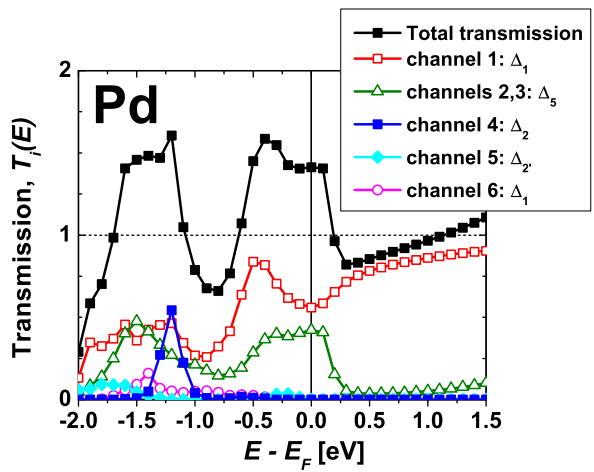}
\caption{(Color online)
Eigenchannel decomposition of the transmission for
the Pd contact shown in Fig.~1.}
\end{center}
\end{figure}

  We turn back to Ni and Co contacts and consider a situation
when a relative orientation
of magnetizations in the leads is antiparallel (AP),
so that an abrupt atomic-scale domain wall is formed as shown in Table~II.
We see from Table~I that, both for Ni and Co, the AP conductance
reflects the structure of the minority spin channels and
consists of a $\Delta_1$ and a $\Delta_5$ channel.
For the atomic configuration shown in Fig.~1,
we obtained "optimistic" MR values: $-17\%$ in case of Ni,
and $+23\%$ in case of Co, which are quite small
in accordance with our previous study.\cite{PRB_MR}
However, the precise MR values as well as transmission curves for Co,
differ from the results reported in our earlier work because of
the different geometrical configurations of atomic contacts. The reason
is that the transmission of $d$-like channels
is quite sensitive to the exact geometry. \cite{PRB_MR}
We mention here, that a more accurate full-potential approach and an
improved description of the electron correlations
for localized $d$-electrons can somewhat affect presented results.
That is also valid for the effects of atomic relaxations which were
neglected. In particular, the exact values for the transmission probabilities
and MR at the Fermi level reported in this study for different
systems could be slightly changed. However, more precise calculations
obviously will not affect the physical results of the present work.

Evident conclusions follow from presented examples. First,
in contrast to earlier studies,\cite{Half_int_FM,Half_int_exp}
the ferromagnetic Ni and Co contacts do not show any tendency to close
one spin channel. On the contrary, both spin channels contribute
to the conductance that gives only moderate magnetoresistance values.
Independent of the geometry of the atomic contact the
minority spin channel will include a sum of fractional contributions
from many modes because the $d$ states are always present at the Fermi level.
That agrees with later experiments by Untiedt {\it et al.},\cite{Untiedt} where
the absence of conductance quantization for ferromagnetic Fe, Co and Ni
contacts was clearly confirmed.

Second, an abrupt, atomic-scale domain wall pinned to the constriction
does not show an impressive MR effect. For a fixed atomic
configuration, the P and AP conductances are of the same order.
Most likely, that more sophisticated calculations, involving relaxation
effects and noncollinear magnetic moments in the domain wall, will not be
able to change this statement.\cite{Tsymbal_PRL,comment}
A recent research\cite{Czerner} towards transport in nanocontacts
with noncollinear moments shows that energetically preferable noncollinear
magnetic order results in a larger domain wall width
as compared to the abrupt, collinear wall considered in the present
paper. That leads to weakened scattering of electrons and a further
reduction of the MR values.

\begin{table}[t]
\caption{Spin magnetic moments (in $\mu_B$) at atoms
forming Ni and Co contacts shown in Fig.~1 for the parallel (P)
and the antiparallel (AP) orientation of magnetizations in the leads.
Bulk magnetic moments are $0.62~\mu_B$ for Ni, and $1.62~\mu_B$ for~Co.}
\begin{ruledtabular}
\begin{tabular}{lcccc}
{} & \multicolumn{2}{c}{Ni} & \multicolumn{2}{c}{Co} \\
\cline{2-3} \cline{4-5}
Atom & P & AP & P & AP \\ \hline
Surface        & $\phantom{+}0.66$ & $\phantom{+}0.66$ & $\phantom{+}1.78$ & $\phantom{+}1.78$ \\
Contact$-1$      & $\phantom{+}0.70$ & $\phantom{+}0.69$ & $\phantom{+}1.83$ & $\phantom{+}1.83$ \\
1st contact  & $\phantom{+}0.68$ & $\phantom{+}0.54$ & $\phantom{+}1.76$ & $\phantom{+}1.64$ \\
2nd contact  & $\phantom{+}0.68$ & $-0.54$ & $\phantom{+}1.76$ & $-1.64$ \\
Contact$+1$    & $\phantom{+}0.70$ & $-0.69$ & $\phantom{+}1.83$ & $-1.83$ \\
Surface        & $\phantom{+}0.66$ & $-0.66$ & $\phantom{+}1.78$ & $-1.78$
\end{tabular}
\end{ruledtabular}
\end{table}

Coming to the experimental situation on ballistic magnetoresistance
(BMR) effect in ferromagnetic contacts,
we point out that large MR values \cite{Huge_MR_1,Hua_Chopra}
were usually measured for much thicker constrictions
(as compared with atomic-sized contacts)
with resistance in the range of hundreds of Ohms.
It is believed,\cite{Magnetostriction1,Magnetostriction2} that
such experiments suffer from many unavoidable artifacts
induced by magnetomechanical effects that mimics the real MR signal
which would come from the spin-polarized transport alone.
However, recent studies by Sullivan {\it et al}.\cite{Sullivan_Ni} and
Chopra {\it et al.}\cite{Chopra_Co} on Ni and Co atomic-sized contacts
report BMR values in the range of $200 \div 2000\%$ with discussion on
the electronic origin of the effect due to domain wall scattering.
Nevertheless attempts to minimize magnetostrictive effects were
undertaken, we just can repeat\cite{Magnetostriction1} that a natural
explanation of these\cite{Sullivan_Ni,Chopra_Co}
and similar experiments \cite{Huge_MR,Viret}
is that, due to magnetization reversal
processes, unstable in time atomic constriction changes its contact area
when magnetic field is applied.
Characteristic steps and jumps in the measured
field-dependent conductance (Fig.~4 of Ref.~\onlinecite{Sullivan_Ni}) or
resistance (Fig.~3a of Ref.~\onlinecite{Chopra_Co},
Fig.~3 of Ref.~\onlinecite{Viret})
are distinct evidence of atomic
reconstructions and fractional changes of the contact cross section.
For example, just eliminating one contact atom from the configuration shown in
Fig.~1 changes conductance of a Ni constriction from $1.2~G_0$ (chain of two atoms, see Table I)
up to $\sim 2.8~ G_0$ (one contact atom only, see Ref.~\onlinecite{Imp_Paper2}),
thus producing $\sim 130$\% MR. Further increase of a contact
area can give arbitrary high MR values, that supports
hypothesis on mechanical nature of the effect.
%\vspace{0.5cm}

\section{Conclusions}
To summarize, we have presented a formalism for the evaluation
of conduction eigenchannels of metallic atomic-sized contacts
from first-principles. We have combined the
{\it ab initio} KKR Green's function approach with the Kubo linear
response theory. Starting from the scattering wave formulation of the
conductance problem, we have built a special representation of the
transmission matrix in terms of local, energy and angular momentum
dependent basis inherent to the KKR method. We have proven
that solutions of the eigenvalue problem for the obtained matrix
are identical to conduction eigenchannels introduced by Landauer and B{\"u}ttiker.
Applications of the method have been presented by studying
ballistic electron transport through Cu, Pd, Ni and Co single-atom contacts.
The symmetry analysis of eigenchannels and its connection to
the orbital classification known from the tight-binding approach
were discussed in detail. Experiments on the electron
transport through magnetic contacts were commented.

\section{Acknowledgement}
We acknowledge a financial support through the
Deutsche Forschungsgemeinschaft (DFG),
Priority Programme 1165: "Nanowires and Nanotubes".

\appendix

\section{}

We consider a practical implementation of the Kubo
formula~(\ref{cond_g}): positions $z=z_{\mathrm{L}}$ and
$z'=z_{\mathrm{R}}$ of the $S_L$ and $S_R$ planes are chosen
within the semi-infinite electrodes, where $\Delta t$-matrix
describing a scattering region is sufficiently small (for details,
see Fig.~2). A~sketch as how to find the conduction eigenchannels
in this case has been given in Sec.~IV.E. Below we complete that
discussion by presenting a mathematical justification of the
method.

We proceed in three steps: (i) since a position of $S_L$ and $S_R$
planes in Eq.~(\ref{cond_g}) does not meet an asymptotic limit
criterion, we expand the Green's function in terms of scattering
Bloch states in contrary to expansion (\ref{G_knL}) involving
unperturbed states; (ii) we use the Kubo formula (\ref{cond_g})
and express conductance as a trace of the appropriately defined
current operator in $\boldsymbol{k}$-space; the obtained operator
is identified with transmission matrix $\tau^{\dagger}\tau$; (iii)
finally, we build up the equivalent site-angular momentum
representation of the transmission matrix, that leads us to the
required formulae [Eqs.~(\ref{g_2terms}), (\ref{TT}) of
Sec.~IV.E].

\subsection{Expansion of the Green's function.}

To begin the proof, we note that the asymptotic representation
(\ref{G_knL}) of the Green's function can be rewritten~as
\begin{eqnarray}
\label{G_pert_WF}
\lefteqn{G^{+}(\mathbf{r},\mathbf{r'},E)|_{z,z'\to \mp \infty}  } \\
\nonumber & = & -2\pi i \sum_{\boldsymbol{kk'}}
\overline\Phi_{\boldsymbol{k}}(\mathbf{r},E)\,
{\left[\tau^{-1}(E)\right]_{\boldsymbol{kk'}}}\
\Phi_{\boldsymbol{k'}}(\mathbf{r'},E)
\end{eqnarray}
where an expansion is performed over perturbed Bloch waves
$\Phi_{\boldsymbol{k}}(\mathbf{r},E) =
\Psi_{\boldsymbol{k}}(\mathbf{r},E)/\sqrt{v_{\boldsymbol{k}}} $
and $\overline\Phi_{\boldsymbol{k}}(\mathbf{r},E) =
\overline\Psi_{\boldsymbol{\kappa}}(\mathbf{r},E)/\sqrt{v_{\boldsymbol{k}}}$
which were introduced in Sec.~III. We remind, that function
$\Phi_{\boldsymbol{k}}(\mathbf{r},E)$ is the solution of the
Lippmann-Schwinger equation (\ref{LS}) associated with the
in-coming unperturbed Bloch wave
$\Psi^{\circ}_{\boldsymbol{k}}(\mathbf{r},E)/\sqrt{v_{\boldsymbol{k}}}$
in the left lead propagating towards atomic constriction
($v_{\boldsymbol{k}} \propto \mathrm{v}^z_{\boldsymbol{k}} > 0$),
while the second function,
$\overline\Phi_{\boldsymbol{k}}(\mathbf{r},E)$, is the solution of
the equation associated with the in-coming unperturbed Bloch wave
$\Psi^{\circ\,*}_{\boldsymbol{k}}(\mathbf{r},E)/\sqrt{v_{\boldsymbol{-k}}}$
in the right lead ($v_{\boldsymbol{-k}} \propto
\mathrm{v}^z_{\boldsymbol{-k}} < 0$).

Consider now a general case, when $z=z_{\mathrm{L}}$ and
$z'=z_{\mathrm{R}}$ points related to planes  $S_L$ and $S_R$ (see
Fig.~2) are not taken infinitely far from atomic constriction. We
note, that for the conductance evaluation one needs only the
back-scattering term $\delta G^{+}(\mathbf{r},\mathbf{r'},E)$ in
Eq.~(\ref{Grr}), which is the solution of the Schr\"odinger
equation without a source term. Therefore, we can expand $\delta
G^{+}(\mathbf{r},\mathbf{r'},E)$ over the eigenfunctions of the
whole system corresponding to energy $E$. These are the
propagating perturbed Bloch states,
\begin{equation}
\label{Psik_pert} \Psi_{\boldsymbol{k}}(\mathbf{r},E) = \sum_{nL}
{\mathscr{C}}_{\boldsymbol{k},nL}(E)\,
\phi_{L}(\mathbf{r}-\mathbf{R}_{n},E),
\end{equation}
and the evanescent states,
\begin{equation}
\label{X_a} {\cal{X}}_{\alpha}(\mathbf{r},E) = \sum_{nL}
{\Gamma}_{\alpha,nL}(E)\, \phi_{L}(\mathbf{r}-\mathbf{R}_{n},E),
\end{equation}
where, in turn, both type of functions are expanded over atomic
orbitals, and ${\mathscr{C}}_{\boldsymbol{k},nL}$ and
${\Gamma}_{\alpha,nL}$ are expansion coefficients.

In particular, matrix ${\mathscr{C}}_{\boldsymbol{\kappa},nL}$ is
related to the matrix ${C}_{\boldsymbol{\kappa},nL}$ corresponding
to the unperturbed Bloch waves (for details, see
Ref.~\onlinecite{pert_BW}):
\[
{\mathscr{C}}_{\boldsymbol{k},nL}(E) = \sum_{n'L'} \left[\bigl\{ 1
- G_0(E)\, \Delta t(E) \bigr\}^{-1}\right]^{nn'}_{LL'}
{C}_{\boldsymbol{k},n'L'}(E)
\]
where the $\Delta t$-matrix describes the whole scattering region
(a vacuum barrier plus an atomic constriction) and $G_0$ is the
structural Green's function of the three-dimensional periodic bulk
crystal.

The evanescent states can be, for example, the surface states
which are perturbed by an atomic constriction, as well as the
states localized around impurities.

Both equations (\ref{Psik_pert}) and (\ref{X_a}) can be joined to
one matrix equation:
\[
\left(
\begin{array}{c}
\Psi(\mathbf{r},E) \\
{\cal{X}}(\mathbf{r},E)
\end{array} \right) =  \left[
\begin{array}{c}
\mathscr{C}(E) \\ \Gamma(E)
\end{array} \right] \phi(\mathbf{r},E),
\]
where (\dots)\ stands for a column-vector, while [\dots]\ denotes
a matrix. We also expand the conjugated states~as
\[
\left(
\begin{array}{c}
\overline\Psi(\mathbf{r},E) \\
{\cal{X}}^{*}(\mathbf{r},E)
\end{array} \right) =  \left[
\begin{array}{c}
\widetilde{\mathscr{C}}(E) \\ \widetilde\Gamma(E)
\end{array} \right] \phi(\mathbf{r},E).
\]

 We expand the back-scattering term $\delta G^{+}(\mathbf{r},\mathbf{r'},E)$
over eigenfunctions $\Psi_{\boldsymbol{k}}(\mathbf{r},E)$ and
${\cal{X}}_{\alpha}(\mathbf{r},E)$ with energy $E$:
\begin{eqnarray}
\nonumber \lefteqn{\delta G^{+}(\mathbf{r},\mathbf{r'},E)
\Bigl|_{\mathbf{r,r'}\in S_{L,R}}} \\
\nonumber & = & \sum_{n\in S_L, n'\in S_R}\sum_{LL'}
\phi^n_L(\mathbf{r},E)\ G^{nn'}_{LL'}(E)\
\phi^{n'}_{L'}(\mathbf{r'},E) \\
& =  & \nonumber \left(
\begin{array}{c}
\overline\Psi(\mathbf{r},E) \\
{\cal{X}}^{*}(\mathbf{r},E)
\end{array} \right)^T
\left[
\begin{array}{cc}
F^{00}(E) & F^{01}(E) \\
F^{10}(E) & F^{11}(E)
\end{array} \right]
\left(
\begin{array}{c}
\Psi(\mathbf{r'},E) \\
{\cal{X}}(\mathbf{r'},E)
\end{array} \right)
\\
\nonumber & = & \phantom{+\ } \sum_{\boldsymbol{k k'}}
\overline\Psi_{\boldsymbol{k}}(\mathbf{r},E)\,
F^{00}_{\boldsymbol{kk'}}(E)\,
\Psi_{\boldsymbol{k'}}(\mathbf{r'},E) \\
\nonumber &  & +\ \sum_{\boldsymbol{k}\beta}
\overline\Psi_{\boldsymbol{k}}(\mathbf{r},E)\,
F^{01}_{\boldsymbol{k}\beta}(E)\,
{\cal{X}}_{\beta}(\mathbf{r'},E) \\
\nonumber &  & +\ \sum_{\alpha\boldsymbol{k'}}
{\cal{X}}^{*}_{\alpha}(\mathbf{r},E)\,
F^{10}_{\alpha\boldsymbol{k'}}(E)\,
\Psi_{\boldsymbol{k'}}(\mathbf{r'},E) \\
\nonumber &  & +\ \sum_{\alpha\beta}
{\cal{X}}^{*}_{\alpha}(\mathbf{r},E)\, F^{11}_{\alpha\beta}(E)\,
{\cal{X}}_{\beta}(\mathbf{r},E) \\
\label{G_rr_E_Fexp} & = & \delta
G_{+}^{(1)}(\mathbf{r},\mathbf{r'},E) + \delta
G_{+}^{(2)}(\mathbf{r},\mathbf{r'},E).
\end{eqnarray}
Only the first term
\begin{equation}
\label{G1_F} \delta G_{+}^{(1)}(\mathbf{r},\mathbf{r'},E) =
\sum_{\boldsymbol{k k'}}
\overline\Psi_{\boldsymbol{k}}(\mathbf{r},E)\,
F^{00}_{\boldsymbol{k k'}}(E)\,
\Psi_{\boldsymbol{k'}}(\mathbf{r'},E)
\end{equation}
exists in the asymptotic case, while other three terms, denoted as
$\delta G_{+}^{(2)}(\mathbf{r},\mathbf{r'},E)$, contain
contributions of the evanescent states which decay within the
leads.

 The unknown matrix $F=\{F^{ij}\}$ in Eq.~(\ref{G_rr_E_Fexp}) satisfies
the equation (below we will skip for simplicity the obvious energy
dependence):
\[
\left[ \widetilde{\mathscr{C}}^T\ \widetilde\Gamma^T \right]
\times \left[
\begin{array}{cc}
F^{00} & F^{01} \\
F^{10} & F^{11}
\end{array} \right]
\times \left[
\begin{array}{c}
{\mathscr{C}} \\ \Gamma
\end{array} \right] =  G_{\mathrm{LR}},
\]
where symbol $T$ denotes transpose, and matrix $G_{\mathrm{LR}}$
is introduced which contains selected matrix elements of the
structural Green's function $G_{LL'}^{nn'}$ with $n \in
S_{\mathrm{L}}$ and $n'\in S_{\mathrm{R}}$.

To find matrices $F^{ij}$, we introduce the block-matrix $\left[
{\cal{B}}^{\dagger}\ \Delta^{\dagger} \right]$ which is a solution
of the following equation (here symbol $\dagger$ denotes Hermitian
conjugate):
\begin{equation}
\label{Eq_CG_BD} \left[
\begin{array}{c}
{\mathscr{C}} \\ \Gamma
\end{array} \right] \times
\left[ {\cal{B}}^{\dagger}\ \Delta^{\dagger} \right] = \left[
\begin{array}{cc}
{\mathscr{C}} {\cal{B}}^{\dagger}& {\mathscr{C}}\Delta^{\dagger} \\
\Gamma {\cal{B}}^{\dagger}  & \Gamma \Delta^{\dagger}
\end{array} \right] =
\left[ \begin{array}{cc} 1 & 0 \\ 0 & 1
\end{array} \right].
\end{equation}
This means, that matrix $\left[ {\cal{B}}^{\dagger}\
\Delta^{\dagger} \right]$ is the inverse or (in general case) the
pseudoinverse \cite{Gantmacher} matrix to the matrix with
${\mathscr{C}}$- and $\Gamma$-blocks.\cite{BD_matrix}.

 The equation similar to Eq.~(\ref{Eq_CG_BD}) exists for other matrices:
\begin{equation}
\label{Eq_CG_BDx} \left[
\begin{array}{c}
\widetilde{\cal{B}}^{*} \\
\widetilde\Delta^{*}
\end{array} \right]
 \times \left[
\widetilde{\mathscr{C}}^T\  \widetilde\Gamma^T
 \right]
= \left[ \begin{array}{cc} \widetilde{\cal{B}}^{*}\,
\widetilde{\mathscr{C}}^T &
\widetilde{\cal{B}}^{*}\, \widetilde{\Gamma}^T \\
\widetilde{\Delta}^{*}\, \widetilde{\mathscr{C}}^T &
\widetilde{\Delta}^{*}\, \widetilde{\Gamma}^T
\end{array} \right] =
\left[ \begin{array}{cc} 1 & 0 \\ 0 & 1
\end{array} \right],
\end{equation}
where symbol $*$ denotes complex conjugate of the matrix elements
only.

 With the use of Eqs.~(\ref{Eq_CG_BD}),~(\ref{Eq_CG_BDx})
we find solution for  $F=\{F^{ij}\}$:
\[
\left[\begin{array}{cc} F^{00} & F^{01} \\ F^{10} & F^{11}
\end{array} \right] =
\left[ \begin{array}{cc} \widetilde{\cal{B}}^{*} G_{\mathrm{LR}}
{\cal{B}}^{\dagger} &
\widetilde{\cal{B}}^{*} G_{\mathrm{LR}} {\Delta}^{\dagger} \\
\widetilde{\Delta}^{*}  G_{\mathrm{LR}} {\cal{B}}^{\dagger} &
\widetilde{\Delta}^{*}  G_{\mathrm{LR}} {\Delta}^{\dagger}
\end{array} \right],
\]
Then expansion coefficients of the $\delta G^{(1)}_{+}$ term read
as
\begin{eqnarray*}
%\label{F00}
\lefteqn{F^{00}_{\boldsymbol{kk'}}(E)} \\
& = & \sum_{n\in S_{\mathrm{L}}, n'\in S_{\mathrm{R}}} \sum_{LL'}
\widetilde{\cal{B}}^{*}_{\boldsymbol{k},nL}(E)
\bigl[G_{\mathrm{LR}}(E)\bigr]^{nn'}_{LL'}
{\cal{B}}^{\dagger}_{n'L',\boldsymbol{k'}}(E).
\end{eqnarray*}
With help of matrix $\tau^{-1}_{\mathrm{LR}}(E)$,
\begin{equation}
\label{tau1_kk} \left[
\tau^{-1}_{\mathrm{\,LR}}(E)\right]_{\boldsymbol{kk'}}
 = -\sqrt{v_{\boldsymbol{\kappa}}}\
\frac{F^{00}_{\boldsymbol{kk'}}(E)}{2\pi i} \
 \sqrt{v_{\boldsymbol{\kappa'}}}
\end{equation}
(with positive velocities under square-roots) we write down
Eq.~(\ref{G1_F}) as
\begin{eqnarray}
\label{G1_tau} \lefteqn{\delta
G_{+}^{(1)}(\mathbf{r},\mathbf{r'},E)
\Bigl|_{\mathbf{r,r'} \in S_{L,R}} } \\
\nonumber & = & -2\pi i \sum_{\boldsymbol{kk'}}
\overline\Phi_{\boldsymbol{k}}(\mathbf{r},E)\,
\left[\tau^{-1}_{\mathrm{\,LR}}(E)\right]_{\boldsymbol{kk'}}
\Phi_{\boldsymbol{k'}}(\mathbf{r'},E).
\end{eqnarray}
Comparing obtained equation with one for the asymptotic limit
[Eq.~(\ref{G_pert_WF})] and taking into account that in fact
$\tau^{-1}_{\mathrm{LR}}(E)$ is independent on the positions of
the L and R planes, we obtain:
\begin{equation}
\label{tau_LR_tau}
 \left[\tau^{-1}_{\mathrm{LR}}(E)\right]_{\boldsymbol{kk'}} =
 \left[\tau^{-1}(E)\right]_{\boldsymbol{kk'}}.
\end{equation}

\subsection{Evaluation of conductance}

When conductance is evaluated, only $\delta
G^{(1)}_{+}(\mathbf{r},\mathbf{r}',E)$ term in expansion
(\ref{G_rr_E_Fexp}) contributes. Other term, $\delta
G^{(2)}_{+}(\mathbf{r},\mathbf{r'},E)$, contains the evanescent
states which decay within the leads and, therefore, have zero
velocities along the current flow. When Eq.~(\ref{G1_tau}) is
inserted into the Kubo formula (\ref{cond_g})
%(\ref{cond_g}),
we obtain:
\begin{equation}
%\nonumber
%g & = &  g_0 \int_{S_L} dS \int_{S_R} dS'\
%\delta G^{(1)}_{+}(\mathbf{r},\mathbf{r'},E_F)%
%\stackrel{\leftrightarrow}{\partial_{z}}\,
%\stackrel{\leftrightarrow}{\partial_{z'}} \\
\label{g_JtJt}
%& & \times \ \delta G^{(1)}_{-}(\mathbf{r'},\mathbf{r},E_F) \\
%\nonumber
g =  g_0 \sum_{\boldsymbol{kk'}} \sum_{\boldsymbol{k_1 k'_1}}
{\cal{J}}^{\mathrm{L}}_{\boldsymbol{k_1 k}}
\left[\frac{1}{\tau_{\mathrm{LR}}}\right]_{\boldsymbol{ k k'}}
{\cal{J}}^{\mathrm{R}}_{\boldsymbol{k'k'_1}}
\left[\frac{1}{\tau^{\dagger}_{\mathrm{RL}}}
\right]_{\boldsymbol{k'_1 k_1}},
\end{equation}
where all matrices are evaluated at the Fermi energy ($E_F =
E_{\boldsymbol{k}}=E_{\boldsymbol{k'}}= \dots$). Here we have
introduced the current operators related to the left (L) lead,
\begin{eqnarray}
\label{JL_def}
\lefteqn{{\cal{J}}^{\mathrm{L}}_{\boldsymbol{k k'}}(E) } \\
\nonumber & = & 2\pi \int\limits_{S_{\mathrm{L}}} dS \left[
\overline\Phi^{\,*}_{\boldsymbol{k}}(\mathbf{r},E)\,
i\!\stackrel{\leftrightarrow}{\partial_{z}}
\overline\Phi_{\boldsymbol{k'}}(\mathbf{r},E)\right] \ = \ \bigl[
\tau^{\dagger}\tau \bigr]_{\boldsymbol{kk'}},
\end{eqnarray}
and to the right (R) one:
\begin{eqnarray}
\label{JR_def}
\lefteqn{{\cal{J}}^{\mathrm{R}}_{\boldsymbol{kk'}}(E)}  \\
\nonumber & = & 2\pi \int\limits_{S_{\mathrm{R}}} dS \left[
\Phi_{\boldsymbol{k}}(\mathbf{r},E)\,
i\!\stackrel{\leftrightarrow}{\partial_{z}}
\Phi^{\,*}_{\boldsymbol{k'}}(\mathbf{r},E)\right] \ = \ \bigl[
\tau\tau^{\dagger}\bigr]_{\boldsymbol{kk'}},
\end{eqnarray}
where basis $\nu$ of eigenchannels (see Sec.~III) was used to
evaluate the matrix elements. With help of (\ref{JL_def}) and
(\ref{JR_def}), equation (\ref{g_JtJt}) for conductance $g$ takes
a form:
\begin{eqnarray}
\label{g_TrJL} g & = &  g_0
 \mathrm{Tr}_{\boldsymbol{k}}
\left( \tau^{\dagger} \tau\, {\cal{O}}^{\mathrm{\,LR}}
\right) \\
\nonumber & = & g_0 \mathrm{Tr}_{\boldsymbol{k}}
\left(\tau^{\dagger} \tau \right)\ =\
 g_0 \mathrm{Tr}_{\boldsymbol{k}}
\left( {\cal{J}}^{\mathrm{L}} \right),
\end{eqnarray}
where ${\cal{O}}^{\mathrm{\,LR}}_{\boldsymbol{kk'}} =
\delta_{\boldsymbol{kk'}}$ is the unitary operator in the
$\boldsymbol{k}$-space:
\begin{eqnarray}
\label{O_kk} {\cal{O}}^{\mathrm{\,LR}} & = &
\tau^{-1}_{\mathrm{LR}}\, {\cal{J}}^{\mathrm{R}}\,
\bigl( \tau^{\dagger}_{\mathrm{RL}}\bigr)^{-1}\\
\nonumber & = &  \tau^{-1}_{\mathrm{LR}} \bigl( \tau\tau^{\dagger}
\bigr) \bigl( \tau^{\dagger}_{\mathrm{RL}}\bigr)^{-1}  = 1,
\end{eqnarray}
and Eqs.~(\ref{tau_LR_tau}) and (\ref{JR_def}) were used.

When the perturbed Bloch waves are expanded over atomic orbitals,
\begin{eqnarray*}
%\label{varPhi_kk}
\Phi_{\boldsymbol{\kappa}}(\mathbf{r},E) & = &
\frac{1}{\sqrt{v_{\boldsymbol{k}}}} \
\Psi_{\boldsymbol{k}}(\mathbf{r},E) \\
& = &  \frac{1}{\sqrt{2\pi}} \sum_{n\in S_R}\sum_{L}
{X}_{\boldsymbol{k},nL}(E)\, \phi^{n}_L(\mathbf{r},E), \\
\overline\Phi_{\boldsymbol{k}}(\mathbf{r},E) & = &
\frac{1}{\sqrt{v_{\boldsymbol{k}}}}\
\overline\Psi_{\boldsymbol{k}}(\mathbf{r},E) \\
& = & \frac{1}{\sqrt{2\pi}} \sum_{n\in S_L}\sum_{L}
\widetilde{X}_{\boldsymbol{k},nL}(E)\, \phi^{n}_L(\mathbf{r},E),
\end{eqnarray*}
current operators ${\cal{J}}^{\mathrm{L}}$ and
${\cal{J}}^{\mathrm{R}}$ take a form (energy dependence is
skipped):
\begin{eqnarray}
\label{JLR_kk} {\cal{J}}^{\mathrm{L}}_{\boldsymbol{kk'}} & = &
\sum_{nn'\in S_L}\sum_{LL'} \widetilde{X}^{*}_{\boldsymbol{k},nL}
\left[D_{\mathrm{L}} \right]^{nn'}_{LL'}
\widetilde{X}^{*\dagger}_{n'L',\boldsymbol{k'}}, \\
\nonumber {\cal{J}}^{\mathrm{R}}_{\boldsymbol{kk'}}  & = &
\sum_{nn'\in S_R}\sum_{LL'} {X}_{\boldsymbol{k},nL}
\left[D_{\mathrm{R}} \right]^{nn'}_{LL'}
{X}^{\dagger}_{n'L',\boldsymbol{k'}},
\end{eqnarray}
where operators $X$ and $\widetilde{X}$ are defined as
\begin{eqnarray}
\label{XC_1} {X}_{\boldsymbol{k},nL}(E) & = &
\left(\frac{2\pi}{v_{\boldsymbol{k}}} \right)^{1/2}
{\mathscr{C}}_{\boldsymbol{k},nL}(E), \\
\nonumber \widetilde{X}_{\boldsymbol{k},nL}(E) & = &
\left(\frac{2\pi}{v_{\boldsymbol{k}}}\right)^{1/2}
\widetilde{\mathscr{C}}_{\boldsymbol{k},nL}(E),
\end{eqnarray}
and operators $D_{\mathrm{L}}$ and $D_{\mathrm{R}}$ were
introduced in Eq.~(\ref{D_LR}).

\subsection{Decomposition of current operator}

 According to Eq.~(\ref{g_TrJL}) conductance is determined
by operator $\cal{J}^{\mathrm{L}} =\tau^{\dagger}\tau$. The formal
representation (\ref{JLR_kk}) of the current operator
$\cal{J}^{\mathrm{L}}$ in $\boldsymbol{k}$-space is not suitable
for practical implementation. It can be used, however, to find an
equivalent representation in $(n,L)$-space.  Using (\ref{JLR_kk})
together with representation of operator $D_{\mathrm{L}}$
[Eqs.~(23), (24)] we can decompose $\cal{J}^{\mathrm{L}}$ in two
terms:
\begin{equation}
\label{JL_2tems} {\cal{J}}^{\mathrm{L}} =
{\cal{J}}^{(+)}_{\mathrm{L}} + {\cal{J}}^{(-)}_{\mathrm{L}},
\end{equation}
where
\begin{equation}
{\cal{J}}^{(+)}_{\mathrm{L}} = + \widetilde{X}^{*}\,
U_{\mathrm{L}} \left( \begin{array}{cc} \Delta^{+}_{\mathrm{L}} &
0 \\ 0 & 0 \end{array} \right)
 U_{\mathrm{L}}^{\dagger}\,
\bigl[\widetilde{X}^{*}\bigr]^{\dagger}
\end{equation}
has only positive (non-negative) eigenvalues, while
\begin{equation}
{\cal{J}}^{(-)}_{\mathrm{L}} = - \widetilde{X}^{*}\,
U_{\mathrm{L}} \left( \begin{array}{cc} 0 & 0 \\ 0 &
\Delta^{-}_{\mathrm{L}}  \end{array} \right)
U_{\mathrm{L}}^{\dagger}\, \bigl[\widetilde{X}^{*}
\bigr]^{\dagger}
\end{equation}
has only negative (or zero-) eigenvalues, and
$\Delta_{\mathrm{L}}^{\pm}$ are diagonal matrices with nonnegative
elements [Eqs.~(23), (24)]. The advantage of operators
$\cal{J}^{(\pm)}_{\mathrm{L}}$ over operator
$\cal{J}_{\mathrm{L}}$ is that matrices
$\Delta^{\pm}_{\mathrm{L}}$ are positive definite, while
anti-symmetric matrix $D_{\mathrm{L}}$ is not. As we show below in
Appendix B, that makes possible to find for operators
${\cal{J}}^{(\pm)}_{\mathrm{L}}$ an equivalent, site-angular
momentum $(n,L)$ representation, while it is not the case for full
operator ${\cal{J}}_{\mathrm{L}}$. We mention also, that in the
asymptotic limit the contribution due to
${\cal{J}}^{(-)}_{\mathrm{L}}$ vanishes. However, in general case,
spectra of both operators should be found.

 To clarify these ideas, let us elucidate a physical meaning of
Eq.~(\ref{JL_2tems}). Consider new basis functions in
$(n,L)$-space which are constructed with help of unitary
matrix~$U_{\mathrm{L}}$,
\[
%\label{chi_alpha}
\chi_{\alpha}(\mathbf{r},E) = \sum_{n\in S_L} \sum_L
\phi^{n}_L(\mathbf{r},E)\,
\bigl[U_{\mathrm{L}}(E)\bigr]_{nL,\alpha},
\]
so that the velocity operator $D_{\mathrm{L}} = U_L
D^{0}_{\mathrm{L}} U^{\dagger}_L$ in $(n,L)$-space introduced in
Eq.~(21) is diagonal in the new basis:
\begin{eqnarray}
\label{DL_delta}
\lefteqn{\bigl[D_{\mathrm{L}}^0(E)\bigr]_{\alpha\beta} }  \\
\nonumber & = &  \int_{S_{\mathrm{L}}} dS \left[
\chi^{*}_{\alpha}(\mathbf{r},E)\,
i\!\stackrel{\leftrightarrow}{\partial_{z}}
\chi_{\beta}(\mathbf{r},E)\right] \ = \ \delta^{\circ}_{\alpha}(E)
\, \delta_{\alpha\beta}.
\end{eqnarray}
Let us examine further the wave function
$\overline\Phi_{\boldsymbol{k}}(\mathbf{r},E)$ which defines the
current operator ${\cal{J}}^{\mathrm{L}}$ [Eq.~(\ref{JL_def})].
This function is the solution of the Lippmann-Schwinger equation
associated with the in-coming unperturbed Bloch wave
$\Phi^{\circ\,*}_{\boldsymbol{\kappa}}(\mathbf{r},E)$ (initial
channel) propagating from the right (R) lead towards the
nanocontact and being scattered on it. Within the left (L) lead
($\mathbf{r} \in S_{\mathrm{L}}$) we have:
\[
\overline\Phi_{\boldsymbol{k}}(\mathbf{r},E)\ =\
\overline\Phi^{\,(+)}_{\boldsymbol{k}}(\mathbf{r},E)\ +\
\overline\Phi^{\,(-)}_{\boldsymbol{k}}(\mathbf{r},E),
\]
where
\begin{equation}
\label{Phi_k_pm}
\overline\Phi^{\,(\pm)}_{\boldsymbol{k}}(\mathbf{r},E) =
\frac{1}{\sqrt{2\pi}} \sum_{\alpha^{\pm}}
\chi_{\alpha}(\mathbf{r},E)\, \bigl(\widetilde{X}^{*}
U_{\mathrm{L}} \bigr)^{\dagger}_{\alpha\boldsymbol{k}},
\end{equation}
here the ($\alpha^{\pm}$)-sums run over the half of the indices of
the basis functions $\chi_{\alpha}(\mathbf{r},E)$, corresponding
either to the "positive" or to the "negative" window of the
spectrum. Thus, the wave function
$\overline\Phi_{\boldsymbol{k}}(\mathbf{r},E)$ in the L lead is a
linear combination of two functions. The first one,
$\overline\Phi^{(+)}_{\boldsymbol{k}}(\mathbf{r},E)$ carries the
flux in the initial direction of the in-coming wave with momentum
$\boldsymbol{-k}$, from the L electrode to the R one. The second
function, $\overline\Phi^{(-)}_{\boldsymbol{k}}(\mathbf{r},E)$,
carries the flux in the direction being opposite to the
propagation direction of the in-coming wave. An example
illustrating this idea for free electrons is shown in Fig.~3.

 One can check, that operator ${\cal{J}}_{\mathrm{L}}^{(+)}$ is defined on functions
$\overline\Phi^{(+)}_{\boldsymbol{k}}(\mathbf{r},E)$ only, while
${\cal{J}}_{\mathrm{L}}^{(-)}$ is defined on
$\overline\Phi^{(-)}_{\boldsymbol{\kappa}}(\mathbf{r},E)$:
\[
\bigl[ {\cal{J}}_{\mathrm{L}}^{(\pm)}(E)\bigr]_{\boldsymbol{kk'}}
= 2\pi \int\limits_{S_L} dS \left[
\overline\Phi^{(\pm)\,*}_{\boldsymbol{k}}(\mathbf{r},E)\,
i\!\stackrel{\leftrightarrow}{\partial_{z}}
\overline\Phi^{(\pm)}_{\boldsymbol{k'}}(\mathbf{r},E)\right],
\]
because, according to Eqs.~(\ref{DL_delta}) and (\ref{Phi_k_pm}),
the cross-terms involving both functions,
$\overline\Phi^{(+)}_{\boldsymbol{k}}(\mathrm{r},E)$ and
$\overline\Phi^{(-)}_{\boldsymbol{k}}(\mathrm{r},E)$, vanish.

Obviously, that if the $S_{\mathrm{L}}$ plane is chosen way behind
the scattering region, the
$\overline\Phi^{(-)}_{\boldsymbol{k}}(\mathbf{r},E)$ turns to zero
and the contribution due to $\cal{J}_{\mathrm{L}}^{(-)}$ vanishes.
In general case, both operators, $\cal{J}^{(+)}_{\mathrm{L}}$ and
$\cal{J}^{(-)}_{\mathrm{L}}$, should be considered.

\subsection{Spectrum of current operators.}

 In this section we show that transmission probability
of the conduction eigenchannel $\nu$ is given by $T_{\nu} =
\tau_{\nu}^{+} - \tau_{\nu}^{-}$, where $\tau_{\nu}^{+}$ and
$(-\tau_{\nu}^{-})$ are positive and negative eigenvalues of
operators ${\cal{J}}^{(+)}_{\mathrm{L}}$ and
${\cal{J}}^{(-)}_{\mathrm{L}}$, respectively.

Assume, that we know matrix $\theta$ (introduced in Sec.III) which
is the unitary transform to the basis of eigenchannels. According
to Ref.~\onlinecite{conjphi} we have:
\begin{eqnarray*}
\overline\Phi_{\nu}(\mathbf{r},E) & = & \sum_{\boldsymbol{k}}
\overline\Phi_{\boldsymbol{k}}(\mathbf{r},E)\,
\theta^{\dagger}_{\boldsymbol{k}\nu}(E) \\
& = & \overline\Phi_{\nu}^{(+)}(\mathbf{r},E) +
\overline\Phi_{\nu}^{(-)}(\mathbf{r},E).
\end{eqnarray*}
Thus, we have decomposed the wave function of each eigenchannel in
two terms: $\overline\Phi_{\nu}^{(\pm)} = \sum_{\boldsymbol{k}}
\overline\Phi^{\,(\pm)}_{\boldsymbol{k}}\,
\theta^{\dagger}_{\boldsymbol{k}\nu}$.

Consider further the transmission matrix $T$ in the basis of
eigenchannels:
\begin{eqnarray*}
T_{\nu\mu} & = &  T_{\nu} \delta_{\nu\mu}\ =\ \left[
\theta ( \tau^{\dagger} \tau ) \theta^{\dagger} \right]_{\nu\mu} \\
& = &  2\pi \int_{S_{\mathrm{L}}} dS\
\left[\overline\Phi_{\nu}^{\,*}(\mathbf{r},E)\
i\!\stackrel{\leftrightarrow}{\partial_{z}}
\overline\Phi_{\mu}(\mathbf{r},E) \right] \\
& = & T^{(+)}_{\nu\mu} + T^{(-)}_{\nu\mu},
\end{eqnarray*}
with
\[
T^{(\pm)}_{\nu\mu} = 2\pi \int_{S_{\mathrm{L}}} dS\
\left[\overline\Phi_{\nu}^{(\pm)\,*}(\mathbf{r},E)\,
i\!\stackrel{\leftrightarrow}{\partial_{z}}
\overline\Phi^{(\pm)}_{\mu}(\mathbf{r},E)\right].
\]
Operators $T^{(\pm)}$ are related to current operators
${\cal{J}}^{(\pm)}_{\mathrm{L}}$ via the unitary transform
$\theta$: $T^{(\pm)}  = \theta {\cal{J}}_{\mathrm{L}}^{(\pm)}
\theta^{\dagger}$.

The unitary transform $\theta$ diagonalizes the full operator
${\cal{J}}^{\mathrm{L}} = {\cal{J}}^{(+)}_{\mathrm{L}} +
{\cal{J}}^{(-)}_{\mathrm{L}}$. Using the representation
(\ref{JLR_kk}) of operators ${\cal{J}}^{(\pm)}_{\mathrm{L}}$ we
can check, that ${\cal{J}}^{(+)}_{\mathrm{L}}$ and
${\cal{J}}^{(-)}_{\mathrm{L}}$ do not commutate. Thus,  the
$T^{(\pm)}$ operators do not have a diagonal form in the basis of
eigenchannels. Therefore, we represent each of these operators as
sum of diagonal and off-diagonal terms:
\begin{eqnarray}
\label{Tpm_full} T^{(-)}_{\nu\mu} & = &
\stackrel{\circ}{T}^{(-)}_{\nu\mu}
\ +\ \delta T^{(-)}_{\nu\mu}, \\
\nonumber T^{(+)}_{\nu\mu} & = &
\stackrel{\circ}{T}^{(+)}_{\nu\mu} \ -\ \delta T^{(-)}_{\nu\mu},
\end{eqnarray}
here $\delta T^{(-)}_{\nu\nu} = 0$, and off-diagonal contributions
are of different signs because the sum of two matrices is the
diagonal matrix $T_{\nu}\delta_{\mu\nu}$. The diagonal terms are
\begin{equation}
\label{T0_diag} \stackrel{\circ}{T}^{(+)}_{\nu\mu} \
 = \ \tau^{+}_{\nu} \delta_{\nu\mu}, \quad
\stackrel{\circ}{T}^{(-)}_{\nu\mu}
 = - \tau^{-}_{\nu} \delta_{\nu\mu},
\end{equation}
where
\[
\tau^{\pm}_{\nu}\ = \ 2\pi \int_{S_{\mathrm{L}}} dS\
\left[\overline\varPhi_{\nu}^{(\pm)\,*}(\mathbf{r},E) \,
i\!\stackrel{\leftrightarrow}{\partial_{z}}
\overline\varPhi^{(\pm)}_{\nu}(\mathbf{r},E)\right].
\]
and transmission probability of the eigenchannel is given~by
$T_{\nu} = \tau^{+}_{\nu} - \tau^{-}_{\nu}$. Here the first term
arises due to all multiple scattering contributions in the
direction of the current, while the second term is due to
scattering contributions in the opposite direction.

In spite of the fact that operators $T^{(+)}$ and $T^{(-)}$ can
not be diagonalized within the unique unitary transform, one can
show that off-diagonal terms $\delta T^{(-)}_{\nu\mu} \ne 0$ are
small. They are determined by functions
$\overline\Phi^{(-)}_{\nu}(\mathbf{r},E)$ with negative velocities
with respect to the initial velocity of the in-coming channel
$\nu$ incident from the right (R) lead. The function
$\overline\Phi^{(-)}_{\nu}(\mathbf{r},E)$ in the left (L) lead is
small as compared with $\overline\Phi^{(+)}_{\nu}(\mathbf{r},E)$.
When the $S_{\mathrm{L}}$ plane is placed far enough from the
surface, the $\overline\Phi^{(-)}_{\nu}(\mathbf{r},E)$ collects
all multiple scattering events in the opposite direction to the
current flow. However, such scattering processes are possible only
due to small inhomogeneities of the potential which is not exactly
the bulk one around~$z_{\mathrm{L}}$ (see Fig.~3). Thus, we can
write down:
\[
\bigl|\overline\Phi^{(-)}_{\nu}(\mathbf{r},E)\bigr| \ \sim  \
\rho\, \bigl|\overline\Phi^{(+)}_{\nu}(\mathbf{r},E)\bigr|,
\]
where parameter $\rho < 1$ has a meaning of reflection amplitude
(see caption of Fig.~3). The $\rho^2$ is of the order of
reflection probability, which (in realistic calculations) is
$\rho^2 \sim \tau^{-}_{\nu}/\tau^{+}_{\nu} \sim 0.1$. Thus,
off-diagonal terms, $\delta T^{(-)}_{\mu\nu}$, contain small
parameter $\lambda \sim \rho^2 \sim 0.1$. Since $\delta
T^{(-)}_{\mu\mu} = 0$, according to the perturbation theory
\cite{QM} the difference between eigenvalues of operators
$T^{(\pm)}_{\mu\nu}$ [Eq.(\ref{Tpm_full})] and
$\stackrel{\circ}{T}^{(\pm)}_{\nu\mu}$ [Eq.(\ref{T0_diag})] {\it
appears only in the second order}, which is $O(\lambda^2) \sim
0.01$. Therefore, the operators ${\cal{J}}^{(\pm)}_{\mathrm{L}}$
provide a feasible way to find $\tau^{(\pm)}_{\nu}$ with good
enough precision.

 To find the spectrum of positive definite
operators ${\cal{J}}_{\mathrm{L}}^{(\pm)}$ we represent them in
the form $\Sigma_{\pm}^{\dagger}\Sigma_{\pm}$, where
$\Sigma_{\pm}$ maps the site-orbital space on the
$\boldsymbol{k}$-space. We can prove (see Appendix~B) that
spectrum of $\Sigma^{\dagger}_{\pm}\Sigma_{\pm}$ is the same as
for operator $\Sigma_{\pm}\Sigma^{\dagger}_{\pm}$ which is defined
in $(n,L)$-space and reads~as
\[
{\cal{T}}^{(\pm)} = \pm \left(\pm
D^{\pm}_{\mathrm{L}}\right)^{1/2}
\bigl[\widetilde{X}^{*}\bigr]^{\dagger}
{\cal{O}}^{\,\mathrm{LR}}\widetilde{X}^{*} \left(\pm
D^{\pm}_{\mathrm{L}}\right)^{1/2}.
\]
Here we have used the unit operator in $\boldsymbol{k}$-space
${\cal{O}}^{\,\mathrm{LR}} = \tau^{-1}_{\mathrm{LR}}\,
{\cal{J}}_{\mathrm{R}}
\bigl(\tau^{\dagger}_{\mathrm{RL}}\bigr)^{-1} =
\delta_{\boldsymbol{kk'}}$ introduced in Eq.~(\ref{O_kk}).
By using equations (\ref{tau1_kk}), (\ref{O_kk}), % (\ref{VV_k}),
(\ref{JLR_kk}), (\ref{XC_1}), we obtain expression for
operator~${\cal{T}}^{(\pm)}$:
\begin{eqnarray*}
\lefteqn{{\cal{T}}^{(\pm)} } \\
&  =  & \pm \left(\pm D^{\pm}_{\mathrm{L}}\right)^{1/2} \bigl[
\widetilde{X}^{*}\bigr]^{\dagger} \left[ \tau^{-1}_{\mathrm{LR}}\,
{\cal{J}}_{\mathrm{R}}
\bigl(\tau^{\dagger}_{\mathrm{RL}}\bigr)^{-1} \right]
\widetilde{X}^{*} \left(\pm D^{\pm}_{\mathrm{L}}\right)^{1/2} \\
& = & \pm \left(\pm D^{+}_{\mathrm{L}}\right)^{1/2}
\bigl[\widetilde{\mathscr{C}}^{*}\bigr]^{\dagger}
\left(\widetilde{\cal{B}}^{*}\, G_{\mathrm{LR}}\,
{\cal{B}}^{\dagger} \right)
\mathscr{C}\, D_{\mathrm{R}}\, {\mathscr{C}}^{\dagger} \\
& & \times \left({\cal{B}}\, G^{\dagger}_{\mathrm{RL}}\,
[\widetilde{\cal{B}}^{*}]^{\dagger} \right)
\widetilde{\mathscr{C}}^{*} \left(\pm D^{\pm}_{\mathrm{L}} \right)^{1/2} \\
& = &  \pm \left(\pm D^{\pm}_{\mathrm{L}}\right)^{1/2}\,
G_{\mathrm{LR}}\, D_{\mathrm{R}} G^{\dagger}_{\mathrm{RL}}
\left(\pm D^{\pm}_{\mathrm{L}}\right)^{1/2},
\end{eqnarray*}
where orthogonality relation\cite{calBC} $ {\cal{B}}^{\dagger}
\mathscr{C} =
 [\widetilde{\cal{B}}^{*}]^{\dagger} \widetilde{\cal{C}}^{*} = 1$
was used. Thus, we come to Eq.~(30) of Sec.~VI.

\section{}

Let us consider operator $\Lambda_{\boldsymbol{kk'}}(E)$ acting in
$\boldsymbol{k}$-space of a following kind:
\[
\Lambda = X \Delta X^{\dagger},
\]
where $\Delta$ is a positive definite hermitian operator acting in
the site-orbital $(n,L)$-space. Let $\Theta_{\nu\boldsymbol{k}}$
be a unitary transform which diagonalizes operator $\Lambda$,
\[
\Theta \Lambda \Theta^{\dagger} = \mbox{diag}\{\Lambda_{\nu}\}
 = \hat \Lambda_0.
\]
Our goal is to find a spectrum of $\Lambda$ with help of
$(n,L)$-space. Following Sec.III, we represent $\Lambda$ in the
form:
\[
\Lambda = \Sigma^{\dagger} \Sigma = X \Delta X^{\dagger},
\]
with $\Sigma = \Delta^{1/2} X^{\dagger}$ and $\Sigma^{\dagger} = X
\Delta^{1/2}$. A square-root from the positive definite hermitian
operator is defined as
\[
\Delta^{1/2} =  ({\Delta}^{1/2})^{\dagger}
 = U \Delta_0^{1/2}\, U^{\dagger},
\]
where the unitary matrix $U$ transforms $\Delta$ to the diagonal
form $\Delta_0 = U^{\dagger}\, \Delta\, U$.

Let us show that a spectrum of operator $\Lambda =
\Sigma^{\dagger} \Sigma$ acting in the $\boldsymbol{k}$-space, is
the same as a spectrum of operator $\Sigma \Sigma^{\dagger}$
acting in the $(n,L)$-space.  Let $\Omega$ be such matrix that
\[
\Theta \Sigma^{\dagger} \Omega^{\dagger} = \Omega\Sigma
\Theta^{\dagger} = \mbox{diag}\{\sqrt{\Lambda_{\nu}}\} = \hat
\lambda_0.
\]
Solution for $\Omega$ is $\Omega_{\nu,nL} = \sum_{\boldsymbol{k}}
\mbox{diag}\{1/\sqrt{\Lambda_{\nu}}\}
\Theta_{\nu\boldsymbol{k}}\Sigma^{\dagger}_{\boldsymbol{k},nL} $.
Matrix $\Omega$ diagonalizes $\Sigma\Sigma^{\dagger}$:
\begin{eqnarray*}
\Omega (\Sigma\Sigma^{\dagger}) \Omega^{\dagger} & = &
\hat\lambda_0^{-1} \Theta \Sigma^{\dagger}\,
(\Sigma\Sigma^{\dagger})\,
\Sigma \Theta^{\dagger}  \hat\lambda^{-1}_0 \\
& = & \hat\lambda_0^{-1}\, ( \Theta \Sigma^{\dagger} \Sigma
\Theta^{\dagger} )\,
(\Theta \Sigma^{\dagger} \Sigma \Theta^{\dagger}) \, \hat\lambda^{-1}_0 \\
& = & \hat\lambda_0^{-1} \, \hat\Lambda_0\, \hat\Lambda_0
\hat\lambda^{-1}_0\ = \
\mbox{diag}\{\Lambda_{\nu}\}. \\
{}
\end{eqnarray*}

To finish the proof, one has to check that $\Omega$ is indeed the
unitary transform, i.e.\ the following properties should hold: (i)
$\Omega\Omega^{\dagger} = \delta_{\nu\nu'}$, and (ii)
$\Omega^{\dagger} \Omega = \delta_{nn'}\delta_{LL'}$. First, we
check property~(i):
\begin{eqnarray*}
\Omega\Omega^{\dagger} & = & (\hat\lambda_0^{-1} \Theta
\Sigma^{\dagger})\,
(\Sigma\Theta^{\dagger}\hat\lambda_0^{-1} ) \\
& = & \hat\lambda_0^{-1}\, (\Theta \Sigma^{\dagger}
\Sigma\Theta^{\dagger})\, \hat\lambda_0^{-1}\ =\
\hat\lambda_0^{-1} \hat\Lambda_0 \hat\lambda_0^{-1} = 1.
\end{eqnarray*}
Next, we check property~(ii):
\begin{eqnarray*}
\Omega^{\dagger}\Omega & = &
(\Sigma\Theta^{\dagger}\hat\lambda_0^{-1} )\,
(\hat\lambda_0^{-1} \Theta \Sigma^{\dagger}) \\
& = & \Sigma\,(\Theta^{\dagger}\hat\Lambda_0^{-1} \Theta )\,
\Sigma^{\dagger} \ =\  \Sigma \bigl[ \Sigma^{\dagger} \Sigma
\bigr]^{-1} \Sigma^{\dagger} = 1,
\end{eqnarray*}
so the proof is complete.


\begin{thebibliography}{99}

\bibitem{STM}
R.\ Wiesendanger. {\it Scanning Probe Microscopy and Spectroscopy.}\
Cambridge University Press, Cambridge (1994).

\bibitem{Muller}
C.\ J.\ Muller, J.\ M.\ van Ruitenbeek, and L.\ J.\ de Jongh,
Physica C {\bf 191}, 485 (1992).

\bibitem{MCBJ}
C.\ J.\ Muller, J.\ M.\ van Ruitenbeek, and L.\ J.\ de Jongh,
Phys.\ Rev.\ Lett. {\bf 69}, 140 (1992);
J.\ M.\ van Ruitenbeek, Naturwissenschaften {\bf 88}, 59 (2001).

\bibitem{Ruitenbeek}
N.~Agra\"\i t, A.~Levy Yeyati and J.~M. van Ruitenbeek,
Physics Reports {\bf 377}, 81 (2003).

\bibitem{Ohnishi}
H.\ Ohnishi, Yu.\ Kondo, and K.\ Takayanagi, Nature {\bf 395}, 780 (1998).

\bibitem{Yanson_Nature_1}
A.\ I.\ Yanson, G.\ Rubio Bollinger, H.\ E.\ van den Brom,
N.\ Agra\"\i t, and J.~M. van Ruitenbeek,  Nature {\bf 395}, 783 (1998).

\bibitem{Cuevas_exp}
J.\ C.\ Cuevas, A.\ Levy Yeyati, A.\ Mart\'in-Rodero, G.\ R.\ Bollinger,
C.\ Untiedt, and N.\ Agra\"\i t, Phys.\ Rev.\ Lett. {\bf 81}, 2990 (1998).

\bibitem{Atom_rearrang}
U.\ Landmann, W.\ D.\ Luedtke, N.\ A.\ Burnham, and R.\ J.\ Colton,
Science {\bf 248}, 454 (1990);
A.\ P.\ Sutton and J.\ B.\ Pethica, J. Phys.: Condens. Matter {\bf 2},
5317 (1990)

\bibitem{Brandbyge}
M.\ Brandbyge, J.\ Schi\o tz, M.\ R.\ S{\o}rensen, P.\ Stoltze,
K.\ W.\ Jacobsen, J.\ K.\ N{\o}rskov,
L.\ Olesen, E.\ Laegsgaard, I.\ Stensgaard, and F.\ Besenbacher,
Phys.\ Rev.\ B {\bf 52}, 8499 (1995).

\bibitem{Todorov}
T.~N. Todorov and A.~P. Sutton, Phys.\ Rev.\ B {\bf 54}, R14234 (1996).

\bibitem{Agrait}
N.\ Agra\"\i t, G.\ Rubio, and S.\ Vieira, Phys.\ Rev.\ Lett. {\bf 74}, 3995 (1995);
G.\ Rubio, N.\ Agra\"\i t, and S.\ Vieira, Phys.\ Rev.\ Lett. {\bf 76}, 2302 (1996).

\bibitem{Landauer}
Ya.\ M.\ Blanter and M.\ B\"uttiker, Physics Reports {\bf 336}, 1 (2000);
M.\ B\"uttiker, Phys.\ Rev.\ B {\bf 46}, 12485 (1992);
M.\ B\"uttiker, Phys.\ Rev.\ Lett.\ {\bf 57}, 1761 (1986);
M.\ B\"uttiker, Y.\ Imry, R.\ Landauer, and S.\ Pinhas,
Phys.\ Rev.\ B {\bf 31}, 6207 (1985).

\bibitem{Scheer}
E.\ Scheer, P.\ Joyez, D.\ Esteve, C.\ Urbina, M.\ H.\ Devoret,
Phys.\ Rev.\ Lett. {\bf 78}, 3535 (1997).

%\bibitem{SGS}
%Subgap structure

\bibitem{MAR}
T.\ M.\ Klapwijk, G.\ E.\ Blonder and M. Tinkham,
Physica B, {\bf 109 - 110}, 1657 (1982).

\bibitem{Cuevas_TB_model}
J.\ C.\ Cuevas, A.\ L.\ Yeyati, and A.\ Mart\'in-Rodero,
Phys.\ Rev.\ Lett. {\bf 80}, 1066 (1998).

\bibitem{Scheer_Nature}
E.\ Scheer, N.\ Agra{\"\i}t, J.\ C.\ Cuevas, A.\ Levy Yeyati,
B.\ Ludoph, A.\ Martin-Rodero, G.\ Rubio Bollinger, J.\ M.\ van Ruitenbeek,
and C.\ Urbina, Nature {\bf 394}, 154 (1998).

\bibitem{FE_models}
J.\ A.\ Torres, J.\ I.\ Pascual, and J.\ J.\ S\'aenz,
Phys.\ Rev.\ B {\bf 49}, 16581 (1994);
J.\ A.\ Torres,  and J.\ J.\ S\'aenz,
Phys.\ Rev.\ Lett. {\bf 77}, 2245 (1996);
A.\ M.\ Bratkovsky and S.\ N.\ Rashkeev, Phys.\ Rev.\ B {\bf 53}, 13074 (1996).

\bibitem{FE_Brandbyge}
M.\ Brandbyge, K.\ W.\ Jacobsen, and J.\ K.\ N{\o}rskov,
Phys.\ Rev.\ B {\bf 55}, 2637 (1997).

\bibitem{Lang}
N. D. Lang, Phys. Rev. B {\bf 52}, 5335 (1995);
N. D. Lang, Phys. Rev. Lett. {\bf 79}, 1357 (1997);
N. D. Lang, Phys. Rev. B {\bf 55}, 9364 (1997);

\bibitem{Kobayashi}
N.\ Kobayashi, M.\ Brandbyge and M.\ Tsukada, Jpn.\ J.\ Appl.\ Phys. {\bf 38}, 336 (1999);
N.\ Kobayashi, M.\ Brandbyge and M.\ Tsukada, Phys.\ Rev.\ B {\bf 62}, 8430 (2000);
N.\ Kobayashi, M.\ Aono, and M.\ Tsukada, Phys.\ Rev.\ B {\bf 64}, 121402(R) (2001);
K.\ Hirose, N.\ Kobayashi, and M.\ Tsukada, Phys.\ Rev.\ B {\bf 69}, 245412 (2004).

\bibitem{TB_models}
M.\ Brandbyge, N.\ Kobayashi, and M.\ Tsukada, Phys.\ Rev.\ B {\bf 60}, 17064 (1999).

\bibitem{TB_models_1}
J.\ C.\ Cuevas, A.\ Levy Yeyati, A.\ Mart\'in-Rodero, G.\ R.\ Bollinger,
C.\ Untiedt, and N.\ Agra{\"\i}t, Phys.\ Rev.\ Lett. {\bf 81}, 2990 (1998);

\bibitem{Ab-initio}
A.\ Nakamura, M.\ Brandbyge, L.\ B.\ Hansen, and K.\ W.\ Jacobsen,
Phys.\ Rev.\ Lett. {\bf 82}, 1538 (1999);
K.\ S.\ Thygesen, M.\ V.\ Bollinger, and K.\ W.\ Jacobsen,
Phys.\ Rev.\ B {\bf 67}, 115404 (2003);
P.\ Jel{\'i}nek, R.\ P{\'e}rez, J.\ Ortega, and F.\ Flores,
Phys.\ Rev.\ B {\bf 68}, 085403 (2003);
Y.\ Fujimoto and K.\ Hirose, Phys.\ Rev.\ B {\bf 67}, 195315 (2003);
K.\ Palot\'as, B.\ Lazarovits, L.\ Szunyogh, and P.\ Weinberger,
Phys.\ Rev.\ B {\bf 70}, 134421 (2004);
P.\ A.\ Khomyakov and G.\ Brocks, Phys. Rev. B {\bf 70}, 195402 (2004).

\bibitem{Ab-initio_1}
M.\ Brandbyge, J.-L.\ Mozos, P.\ Ordej\'on, J.\ Taylor, and K.\ Stokbro,
Phys.\ Rev.\ B {\bf 65}, 165401 (2002);
J.\ Taylor, H.\ Guo, and J.\ Wang, Phys.\ Rev.\ B {\bf 63}, 245407 (2001);
H.\ Mehrez, A.\ Wlasenko, B.\ Larade, J.\ Taylor, and P.\ Gr\"utter, H.\ Guo,
Phys.\ Rev.\ B {\bf 65}, 195419 (2002).

\bibitem{Rocha}
A.\ R.\ Rocha, V.\ M.\ Garc{\'\i}a-Su{\'a}rez,
S.\ Bailey, C.\ Lambert, J.\ Ferrer, and S.\ Sanvito,
Phys.\ Rev.\ B {\bf 73}, 085414 (2006).

\bibitem{new_methods}
A.\ Pecchia  and A.\ Di Carlo, Rep.\ Prog.\ Phys.\ {\bf 67}, 1497 (2004);
G.\ C.\ Solomon, A.\ Gagliardi, A.\ Pecchia,
Th.\ Frauenheim, A.\ Di Carlo, J.\ R. Reimersa, and N.\ S.\ Hush,
J.\ Chem.\ Phys., {\bf 125}, 184702 (2006);
S.\ Kurth, G.\ Stefanucci, C.-O.\ Almbladh, A.\ Rubio, and E.\ K.\ U.\ Gross,
Phys.\ Rev.\ B {\bf 72}, 035308 (2005);
C.\ Verdozzi, G.\ Stefanucci, and C.-O.\ Almbladh,
Phys.\ Rev.\ Lett.\ {\bf 97}, 046603 (2006).

\bibitem{SKKR}
R.\ Zeller, P.\ H.\ Dederichs, B.\ Ujfalussy, L.\ Szunyogh and P.\ Weinberger,
Phys. Rev. B {\bf 52}, 8807 (1995);
R. Zeller, Phys. Rev. B {\bf 55}, 9400 (1997).
N.\ Papanikolaou, R.\ Zeller, and P.\ H.\ Dederichs,
J.\ Phys.: Condens.\ Matter {\bf 14}, 2799 (2002).

%\bibitem{SKKR_review}
%N.\ Papanikolaou, R.\ Zeller, and P.\ H.\ Dederichs,
%J.\ Phys.: Condens.\ Matter {\bf 14}, 2799 (2002).

\bibitem{BarStone}
H.\ U.\ Baranger and A.\ D.\ Stone, Phys.\ Rev.\ B {\bf 40}, 8169 (1989).

\bibitem{Imp_Paper1}
N.\ Papanikolaou, A.\ Bagrets, and I.\ Mertig, J.\ Phys.: Conf. Ser.\
{\bf 10}, 109 (2005).

\bibitem{Imp_Paper2}
A.\ Bagrets, N.\ Papanikolaou, and I.\ Mertig,
Phys.\ Rev.\ B {\bf 73}, 045428 (2006).

%\bibitem{Appendix_EPAPS}
%See EPAPS Document No. E-PRBMDO-XX-XXXXX for Appendices~A and~B where
%we present a complete mathematical proof on the validity of the procedure
%for the evaluation of eigenchannels briefly outlined in Sec.IV.E. This
%document can be reached via a direct link in the online article's HTML
%reference section or via the EPAPS homepage
%(http://www.aip.org/pubservs/epaps.html).

\bibitem{Mavropoulos}
Ph.\ Mavropoulos, N.\ Papanikolaou, and P.\ H.\ Dederichs,
Phys.\ Rev.\ B, {\bf 69}, 125104 (2004).

\bibitem{Ludoph}
B.\ Ludoph and J.\ M.\ van Ruitenbeek,
Phys.\ Rev. B {\bf 61}, 2273 (2000).

\bibitem{Half_int_exp}
T.\ Ono, Y.\ Ooka, H.\ Miyajima, and Y.\ Otani, Appl.\ Phys.\ Lett.\
{\bf 75}, 1622 (1999); V.\ Rodrigues, J.\ Bettini, P.\ C. Silva, and D.\ Ugarte,
Phys.\ Rev.\ Lett.\ {\bf 91}, 096801 (2003).

\bibitem{Viret}
M.\ Viret, S.\ Berger, M.\ Gabureac, F.\ Ott, D.\ Olligs, I.\ Petej,
J.\ F.\ Gregg, C.\ Fermon, G.\ Francinet, and G.\ Le Goff,
Phys.\ Rev.\ B {\bf 66}, 220401(R) (2002).

\bibitem{PRB_MR}
A.\ Bagrets, N.\ Papanikolaou, and I.\ Mertig,
Phys.\ Rev.\ B {\bf 70}, 064410 (2004).

\bibitem{TB-LMTO-Ni}
A.\ K.\ Solanki, R.\ F.\ Sabiryanov, E.\ Y.\ Tsymbal, S.\ S.\  Jaswal,
Jour.\ Magn.\ Magn.\ Mater.\ {\bf 272-276}, 1730 (2004).

\bibitem{Ni_Spain}
D.\ Jacob, J.\ Fern{\'a}ndez-Rossier, and J.\ J.\ Palacios,
Phys.\ Rev.\ B {\bf 71}, 220403(R) (2005).

\bibitem{Pt_Spain}
J.\ Fern{\'a}ndez-Rossier, D.\ Jacob, C.\ Untiedt,
and J.\ J.\ Palacios, Phys. Rev. B {\bf 72}, 224418 (2005).

\bibitem{Ni_Smogunov}
A. Smogunov, A.\ Dal Corso, and E. Tosatti,
Phys.\ Rev.\ B {\bf 73}, 075418 (2006).

\bibitem{Huge_MR}
N.\ Garc\'ia, M. Mu\~noz, and Y.-W.\ Zhao, Phys.\ Rev.\ Lett.\ {\bf 82}, 2923 (1999).

\bibitem{Huge_MR_1}
N.\ Garcia,  M. Mu\~noz, G.\ G.\ Qian, H.\ Rohrer, I.\ G.\ Saveliev, and Y.-W.\ Zhao,
Appl.\ Phys.\ Lett.\ {\bf 79}, 4550 (2001)

\bibitem{Hua_Chopra}
H.\ D.\ Chopra and S.\ Z.\ Hua, Phys.\ Rev.\ B {\bf 66}, 020403(R) (2002).

\bibitem{Sullivan_Ni}
M.\ R.\ Sullivan, D.\ A.\ Boehm, D.\ A.\ Ateya, S.\ Z. Hua, and H.\ D.\ Chopra,
Phys.\ Rev.\ B\ {\bf 71}, 024412 (2005)

\bibitem{Chopra_Co}
H.\ D.\ Chopra, M.\ R.\ Sullivan, J.\ N.\ Armstrong, and S.\ Z.\ Hua,
Nature Materials {\bf 4}, 832 (2005)

\bibitem{2D_SKKR}
I.\ Turek, V.\ Drchal, J.\ Kudrnovsk{\'y}, M. {\^S}ob,
and P.\ Weinberger, {\it Electronic Structure of Disoredered Alloys, Surafces
and Interfaces} (Kluwer Academic, Boston, 1997).

\bibitem{Brandbyge1}
M.\ Brandbyge, M.\ R.\  S{\o}rensen, and K.\ W.\ Jacobsen,
Phys.\ Rev.\ B {\bf 56}, 14956 (1997).

\bibitem{conjphi}
We note, that a similar result holds for the eigenfunctions
$\overline\Phi_{\nu}({\bf r},E)$ of the system defined
as superposition of scattered states coming from the right lead:
$
%\label{psi_nu}
\overline\Phi_{\nu}(\mathbf{r},E) = \sum_{\boldsymbol{k}}
\overline{\Phi}_{\boldsymbol{k}}(\mathbf{r},E)\,
\theta^{\dagger}_{\boldsymbol{k}\nu}(E)
$,
where the perturbed Bloch state
$\overline{\Phi}_{\boldsymbol{k}}(\mathbf{r},E)$
is the solution of Eq.~(\ref{LS})
corresponding to the initial in-coming state
${\Phi}^{\circ *\,\mathrm{in}}_{\boldsymbol{k}}(\mathbf{r},E)$
in R ($z \to +\infty$).  The notation with bar,
$\overline\Phi_{\nu}({\bf r},E)$, is used to distinguish
from $\Phi_{\nu}({\bf r},E)$.
Taking into account the property of microscopic
reversibility for the transmission amplitudes,
$\tau_{\boldsymbol{-\kappa;
-\kappa'}} = \tau_{\boldsymbol{\kappa'\kappa}}$,
and Eqs.~(\ref{theta}) and (\ref{Curr_Bloch1}),
we obtain:
\[
%\label{Curr_Left}
\int_{S_{\mathrm{L}}} dS \left[
\overline\Phi_{\nu}(\mathbf{r},E)\,
i\!\stackrel{\leftrightarrow}%
{\partial_z} \overline{\Phi}^{*}_{\mu}(\mathbf{r},E)
\right]_{z\to -\infty} =
-\, \frac{T_{\nu}(E)}{2\pi}\, \delta_{\nu\mu}.
\]

\bibitem{RL_functions}
Since cell potential $V_n(\mathbf{r})$ is real,
both regular $R^n_L({\mathbf{r}},E)$ and irregular $H^n_L({\mathbf{r}},E)$
solutions of the radial Schr{\"o}dinger equation can be found as real valued functions.
However, in practical implementation of the KKR method, we find $R_L({\mathbf{r}},E)$
as solution of the Lippmann-Schwinger equation for the incoming
spherical Bessel function $j_l(\sqrt{E}r)Y_L(\mathbf{\hat{r}})$, where
$Y_L(\mathbf{\hat{r}})$ is a real spherical harmonic.
For example, in case of spherical potentials (ASA) employed in this work,
such defined functions $R_L({\mathbf{r}},E)$ carry multipliers $\exp[i\eta^n_l(E)]$,
where $\eta^n_l(E)$ is a scattering phase shift. Without loss of generality,
these phase factors can be ascribed to structure
constants $G^{nn'}_{LL'}(E)$ [see Eq.~(\ref{Grr})]
and solutions $R^n_L({\mathbf{r}},E)$ can be considered as real valued functions.

\bibitem{KKR_band_str}
W.\ Kohn and N.\ Rostoker, Phys. Rev. {\bf 94}, 1111 (1954);
B.\ Segall, Phys. Rev. {\bf 105}, 108 (1957);
F.\ S.\ Ham and B.\ Segall, Phys. Rev. {\bf 124}, 1786 (1961).

\bibitem{B_matrix}
This expression reads
$B_{nL,\boldsymbol{k}}^{\dagger}(E) = \sum_{L'}
\Lambda_{LL'}(E;E_{\boldsymbol{k}}) $
$\times \left[{1}/{N_z}\, e^{-i\mathbf{k}\mathbf{R}_n}
C^{\circ\dagger}_{L'\boldsymbol{k}}(E_{\boldsymbol{k}})\right]$,
where a matrix $\Lambda$ is introduced:
$\Lambda_{LL'}(E;E') = (1/{V_0}) \int_{V_0}
d^3\mathbf{r} \left[\phi^n_{L}(\mathbf{r},E)\,
{\phi}^{n}_{L'}(\mathbf{r},E')\right]$,
here $V_0 = V/N$ is a volume of a unit cell,
whereas $N = N_x N_y N_z$ is number of atoms in a supercell.
Because of orthogonality of the basis functions,
$\int_V d^3\mathbf{r} \left[R_{L}(\mathbf{r},E)\,
R_{L'}(\mathbf{r},E')\right] \sim \delta_{LL'}\,
\delta(E-E')$, the major contribution to Eq.~(\ref{phi_nL})
comes from ${\boldsymbol{k'}}$-states with
$E_{\boldsymbol{k'}} \approx E$. This property is used to
prove the second relation in Eq.~(\ref{BC}).


\bibitem{KKR_velocity}
W.\ H.\ Butler,  Phys. Rev. B {\bf 31}, 3260 (1985).

\bibitem{Gantmacher}
F.~R.~Gantmacher. {\it The Theory of Matrices Vol. 1,2},
AMS, 1998.

\bibitem{Krans}
J.\ M.\ Krans, J.\ M.\ van Ruitenbeek, V.\ V.\ Fisun, I.\ K.\ Yanson, and L.\ J.\ de Jongh,
Nature {\bf 375}, 767 (1995).

\bibitem{Exp_noble_metals}
J.\ L.\ Costa-Kr{\"a}mer, N.\ Garc{\'i}a, P.\ Garc{\'i}a-Mochales,
P.\ A.\ Serena, M.\ I.\ Marqu{\'e}s, and A.\ Correia,
Phys.\ Rev. B {\bf 55}, 5416 (1997);

\bibitem{Enomoto}
A.\ Enomoto, S.\ Kurokawa, and A.\ Sakai,  Phys.\ Rev. B {\bf 65}, 125410 (2002).

\bibitem{Yanson_PhD}
A.\ I.\ Yanson, PhD thesis, Universiteit Leiden, The Netherlands (2001)

%\bibitem{DFT_noble_metals}
%DFT based calculations on noble metals contacts ...

\bibitem{Yanson_Nature}
A.\ I.\ Yanson, I.\ K.\ Yanson, and J.\ M.\ van Ruitenbeek,
Nature {\bf 400}, 144 (1999).

\bibitem{Half_int_FM}
H.\ Imamura, N.\ Kobayashi, S.\ Takahashi, and
S.\ Maekawa, Phys.\ Rev.\ Lett.\ {\bf 84}, 1003 (2000).

\bibitem{Untiedt}
C.\ Untiedt, D.\ M.\ T.\ Dekker, D.\ Djukic, and J.\ M.\ van Ruitenbeek,
Phys.\ Rev.\ B {\bf 69}, 081401(R) (2004).

\bibitem{Electr_origin}
G. Tatara, Y.-W. Zhao, M. Mu{\~n}oz,
and N. Garc\'ia, Phys.\ Rev.\ Lett.\ {\bf 83}, 2030 (1999).

\bibitem{Tagirov}
L.\ R.\ Tagirov, B.\ P.\ Vodopyanov, and K.\ B.\ Efetov,
Phys.\ Rev.\ B\ {\bf 65}, 214419 (2002).

\bibitem{NiO_surface}
N.\ Papanikolaou,
J.\ Phys.:\ Condens.\ Matter\ {\bf 15} (2003) 5049

\bibitem{Velev_Butler}
J.\ Velev and W.\ H.\ Butler, Phys.\ Rev.\ B {\bf 69}, 094425 (2004).

\bibitem{NiO_Spain}
D.\ Jacob, J.\ Fern{\'a}ndez-Rossier, and J.\ J.\ Palacios,
Phys.\ Rev.\ B {\bf 74}, 081402(R) (2006)

\bibitem{Magnetostriction1}
M.\ Gabureac, M.\ Viret, F.\ Ott, and C.\ Fermon,
Phys.\ Rev.\ B {\bf 69}, 100401(R) (2004);

\bibitem{Magnetostriction2}
W.\ F.\ Egelhoff, Jr., L.\ Gan, H.\ Ettedgui, Y.\ Kadmon,
C.\ J.\ Powell, P.\ J.\ Chen, A.\ J.\ Shapiro, R.\ D.\ McMichael,
J.\ J.\ Mallett, T.\ P.\ Moffat, M.\ D.\ Stiles,
and E.\ B.\ Svedberg,
J.\ Appl.\ Phys.\ {\bf 95}, 7554 (2004).

\bibitem{Marrows}
C.\ H.\ Marrows, Adv.\ Phys.\ {\bf 54}, 585 (2005).

\bibitem{Chains}
S.\ R.\ Bahn and K.\ W.\ Jacobsen,
Phys.\ Rev.\ Lett.\ {\bf 87}, 266101 (2001).

\bibitem{Vosko_Wilk_Nusair}
S.\ H.\ Vosko, L.\ Wilk, and N.\ Nusair, Can.\ J.\ Phys.\ {\bf 58}, 1200 (1980).

\bibitem{SRA}
D.\ D.\ Koelling and B.\ N.\ Harmon,
J.\ Phys.\ C: Solid State Phys.\ {\bf 10}, 3107 (1977).

\bibitem{Wigner}
L.\ P.\ Bouckaert, R.\ Smoluchowski, and E.\ Wigner,
Phys.\ Rev.\ {\bf 50}, 58 (1936)

\bibitem{2DEG}
B.\ J.\ van Wees, H.\ van Houten, C.\ W.\ J.\ Beenakker,
J.\ G.\ Williamson, L.\ P.\ Kouwenhoven,  D.\ van der Marel,
and C.\ T.\ Foxon, Phys.\ Rev.\ Lett.\ {\bf 60}, 848 (1988)

\bibitem{Pd_wires}
A.\ Delin, E.\ Tosatti, and R.\ Weht, Phys.\ Rev.\ Lett.\ {\bf 92}, 057201 (2004).

\bibitem{Pd_hist}
Sz.\ Csonka, A.\ Halbritter, G.\ Mih{\'a}ly,
O.\ I.\ Shklyarevskii, S.\ Speller, and H.\ van Kempen,
Phys.\ Rev.\ Lett.\ {\bf 93}, 016802 (2004).

\bibitem{Tsymbal_PRL}
J.\ D.\ Burton, R.\ F.\ Sabirianov, S.\ S.\ Jaswal, E.\ Y.\ Tsymbal, and O.\ N.\ Mryasov,
Phys.\ Rev.\ Lett.\ {\bf 97}, 077204 (2006)

\bibitem{comment}
Recent {\it ab intio} calculations by Burton {\it et al.}
(see above Ref.~\onlinecite{Tsymbal_PRL})
predict 340\% domain wall MR in case of one-dimensional, fcc $5\times4$,
Ni wire. This result follows from the quantized conductance being $14~e^2/h$ in case
of uniformly magnetized wire reduced down to $3.2~e^2/h$ in the presence of domain
wall. However, one has to take into account that transmission probabilities of many
wire channels even without domain wall will be significantly reduced
(especially those which are built from $d$ electrons) when realistic
geometry of a contact is considered. That can diminish MR to quite
moderate values comparable to ones estimated in the present work.

\bibitem{Czerner}
M.\ Czerner, B.\ Yavorsky, and I.\ Mertig.
{\it Programme and Abstracts, $\Psi_k$ 2005 Conference},
Schw{\"a}bisch Gm{\"u}nd, Germany, 17-21 September, 2005; p. 408.

\bibitem{pert_BW}
I.\ Mertig, Rep.\ Prog.\ Phys.\ {\bf 62}, 237 (1999).

%\bibitem{Gantmacher}
%F.~R.~Gantmacher. {\it The Theory of Matrices Vol. 1,2}, AMS,
%1998.

\bibitem{BD_matrix}
Without a loss of generality, the number $N_{\boldsymbol{k}}$ of
$\boldsymbol{k}$-points on the isoenergetic surface
$E=\mathrm{const}$ plus number of (possibly existing) evanescent
states (with energy $E$) $N_{\mathrm{evan}}$ can be chosen to be
$N_{\boldsymbol{k}} + N_{\mathrm{evan}} = N_{\mathrm{at}}
L_{\mathrm{max}}$ where $N_{\mathrm{at}} = N_x N_y$ is number of
atoms in the L (or R) atomic planes, so that the inverse matrix
operation is well defined. Even if $N_{\boldsymbol{k}} +
N_{\mathrm{evan}}$ is not exactly equal to $N_{\mathrm{at}}
L_{\mathrm{max}}$ the rectangular matrix $\left[
{\cal{B}}^{\dagger}\ \Delta^{\dagger} \right]$ can be found as the
pseudoinverse matrix (Ref.~\onlinecite{Gantmacher}) to the matrix
with ${\mathscr{C}}$- and $\Gamma$-blocks.

\bibitem{QM}
L. D. Landau and I. M. Lifshitz, {\it Quantum Mechanics, Course of
Theoretical Physics}, Vol. 3 (Pergamon, Oxford, 1977).

\bibitem{calBC}
From Eq.~(\ref{Eq_CG_BD}) we obtain: $ \sum_{n\in S_L}\sum_L
{\mathscr{C}}_{\boldsymbol{k},nL}
{\cal{B}}^{\dagger}_{nL,\boldsymbol{k'}} =
\delta_{\boldsymbol{kk'}} $, and $ \sum_{n\in S_L}\sum_L
{\Gamma}_{\alpha,nL} {\Delta}^{\dagger}_{nL,\beta} =
\delta_{\alpha\beta} $. In order the $n$-sum be converged, matrix
elements of ${\mathscr{C}}$, ${\cal{B}}^{\dagger}$, $\Gamma$ and
$\Delta^{\dagger}$ should carry coefficients $\sim 1/\sqrt{N_s}$
(where $N_s = N_x N_y$ is number of atoms per cross-section of the
Born-von K{\'a}rm{\'a}n supercell). On the other hand, since
matrix $\left[{\cal{B}}^{\dagger}\ \Delta^{\dagger}\right]$ is the
inverse to the one with ${\mathscr{C}}$ and $\Gamma$ blocks, the
following relation holds: $\sum_{\boldsymbol{k}}
{\cal{B}}^{\dagger}_{nL,\boldsymbol{k}}
{\mathscr{C}}_{\boldsymbol{k},nL} + \sum_{\alpha}
\Delta^{\dagger}_{nL,\alpha} \Gamma_{\alpha,nL}  = \delta_{nn'}
\delta_{LL'} $. The second term in this equation is zero, because
the number of evanescent (surface) states $N_{\alpha}$ at given
energy $E$ is much smaller than the total number of surface states
$\sim N_x N_y =  N_s$, so that $ \sum_{\alpha}
\Delta^{\dagger}_{nL,\alpha} \Gamma_{\alpha,n'L'} \sim
{N_{\alpha}}/{N_s} \to 0$. Finally, we obtain: $
\sum_{\boldsymbol{k}} {\cal{B}}^{\dagger}_{nL,\boldsymbol{k}}
{\mathscr{C}}_{\boldsymbol{k},nL} = \delta_{nn'} \delta_{LL'}$.

\end{thebibliography}
\end{document}